\documentclass[11pt,a4paper]{article} \pdfoutput=1 \usepackage{epsfig}
\usepackage{jheppub}

\usepackage[english]{babel}
\usepackage{graphics,amssymb,float}
\usepackage{graphicx}
\usepackage{color}
\usepackage{xspace}
\usepackage{placeins}
\usepackage{rotating}
\usepackage{bm}
\usepackage{amsmath}
\usepackage{indentfirst} 
\usepackage{longtable,multirow}
\usepackage{slashed}
\usepackage{hyperref}
\usepackage{psfrag}
\usepackage[normalem]{ulem}     

\newcommand{\be}{\begin{equation}}
\newcommand{\ee}{\end{equation}}
\newcommand{\bea}{\begin{eqnarray}}
\newcommand{\eea}{\end{eqnarray}}

\newcommand{\ba}{\begin{array}}
\newcommand{\ea}{\end{array}}
\newcommand{\balg}{\begin{align}}
\newcommand{\ealg}{\end{align}}

\newcommand{\mdm}{m_{DM}}

\newcommand{\To}{\Rightarrow}
\newcommand{\dmu}{\partial_{\mu}}
\newcommand{\dMu}{\partial^{\mu}}
\newcommand{\Dmu}{D_{\mu}}
\newcommand{\DMu}{D^{\mu}}

\newcommand{\mH}{\mathcal{H}}
\newcommand{\mG}{\mathcal{G}}

\newcommand{\sigv}{\langle \sigma_{\text{ann}}.v \rangle}
\newcommand{\RNum}[1]{\uppercase\expandafter{\romannumeral #1\relax}}

\def\lsim{\mathrel{\rlap{\lower4pt\hbox{\hskip1pt$\sim$}}
    \raise1pt\hbox{$<$}}}                
\def\gsim{\mathrel{\rlap{\lower4pt\hbox{\hskip1pt$\sim$}}
    \raise1pt\hbox{$>$}}}                

\title{Dark Matter Constraints on Composite Higgs Models}

\author[a,b]{Nayara Fonseca,}
\author[a]{Renata Zukanovich Funchal,}
\author[a]{Andre Lessa,}
\author[c]{Laura Lopez-Honorez}

\emailAdd{nayara@if.usp.br}
\emailAdd{zukanov@if.usp.br}
\emailAdd{lessa@if.usp.br}
\emailAdd{llopezho@vub.ac.be}

\affiliation[a]{
Instituto de F\'{i}sica, Universidade de S\~{a}o Paulo,
\\ R. do Mat\~{a}o 187, S\~{a}o Paulo, SP 05508-900, Brazil }

\affiliation[b]{Center for Theoretical Physics, Massachusetts Institute of Technology, \\
Cambridge, MA 02139, USA}
\affiliation[c]{Vrije Universiteit Brussel and The International Solvay Institutes, \\
Pleinlaan 2, B-1050 Brussels, Belgium}

\abstract{In composite Higgs models the pseudo-Nambu-Goldstone Boson
  (pNGB) nature of the Higgs field is an interesting alternative for
  explaining the smallness of the electroweak scale with respect to the
  beyond the Standard Model scale. In non-minimal models additional
  pNGB states are present and can be a Dark Matter (DM) candidate, if
  there is an approximate symmetry suppressing their decay.  Here we
  assume that the low energy effective theory (for scales much below
  the compositeness scale) corresponds to the Standard Model with a
  pNGB Higgs doublet and a pNGB DM multiplet. We derive general
  effective DM Lagrangians for several possible DM representations
  (under the SM gauge group), including the singlet, doublet and
  triplet cases. Within this framework we discuss how the DM
  observables (relic abundance, direct and indirect detection)
  constrain the dimension-6 operators induced by the strong sector
   assuming that DM behaves as a Weakly Interacting Particle (WIMP)
    and that the relic abundance is settled through the freeze-out
    mechanism. We also apply our general results to two specific
  cosets: $SO(6)/SO(5)$ and $SO(6)/SO(4) \times SO(2)$, which contain
  a singlet and doublet DM candidate, respectively.  In particular we
  show that if compositeness is a solution to the little hierarchy
  problem, representations larger than the triplet are strongly
  disfavored. Furthermore, we find that composite models can have
  viable DM candidates with much smaller direct detection
  cross-sections than their non-composite counterparts, making DM
  detection much more challenging.}

\preprint{MIT-CTP {4623}}

\begin{document}
 \setcounter{tocdepth}{2}
\maketitle

\section{Introduction}

Despite being an enormous triumph to the Standard Model (SM), the
discovery of the Higgs boson \cite{CMSHiggs:2012, AtlasHiggs:2012, Englert:1964, Higgs1:1964, Higgs2:1964} has established the question of how the
electroweak scale is stabilized under the large corrections from new
physics at ultra-violet (UV) scales. The quadratic sensitivity of the
Higgs mass to UV physics, also known as the hierarchy problem, has
been one of the leading motivations for searches of new physics at the
LHC. One of the most investigated solutions to the hierarchy problem
is Supersymmetry. Besides providing a mechanism for stabilizing the
electroweak (EW) scale, the Minimal Supersymmetric Standard Model (for
reviews see \cite{Baer:2006rs,Drees:2004jm,Martin:1997ns}) or MSSM
also has several attractive features, such as gauge coupling
unification and viable dark matter candidates.  However, current LHC
searches and the measured Higgs mass impose severe constraints on the
MSSM.

Composite Higgs models \cite{Kaplan1:1983, Kaplan2:1983, Banks:1984, Georgi1:1984, Georgi2:1984, Dugan:1984} 
are also an interesting solution to the
hierarchy problem. Unlike the MSSM, where the Higgs boson is a
fundamental scalar, in composite models the Higgs doublet is a
pseudo-Nambu-Goldstone Boson (pNGB) appearing in the low energy theory
as a result of the spontaneous breaking of a global symmetry ($\mG \to
\mH$) by a new strong sector dynamics.  In analogy to the pions in
QCD, the Higgs doublet only has derivative couplings and is exactly
massless, except for corrections due to the explicit breaking of
$\mG$.  In the Minimal Composite Higgs Model (MCHM) \cite{MCHM:05},
based on the coset $SO(5)/SO(4)$, the four pNGBs correspond to the
complex Higgs doublet. Nonetheless, other cosets are possible and may
contain a higher number of pNGBs degrees-of-freedom.  It is then an
interesting question whether one of the additional pseudo
Nambu-Goldstone bosons may explain the observed dark matter component
of the universe.  Since this requires at least five pNGBs (the Higgs
doublet plus one DM state), one must consider extensions of the
MCHM. The simplest scenario corresponds to the $SO(6)/SO(5)$ coset
\cite{Frigerio:2012uc, UrbanoCDM:14}, which provides exactly one
additional pNGB. For a  model based on the coset $SO(7)/G2$, see e.g. \cite{Chala:2012}.

Possible explanations to the DM question can now be tested by an
impressive number of experiments, possibly bringing us at the edge of
discovery. Four complementary approaches are followed to identify DM
as a new WIMP. {\it Direct searches} are looking for DM scattering off
heavy nuclei in underground detectors. For the mass ranges that we are
interested in, LUX is for the time being the most sensitive
experiment~\cite{Akerib:2013tjd}, reaching scattering cross-sections
of the order of $10^{-45}$cm$^2$ (for masses around 30 GeV). In this
field, the next data releases are expected for 2015 (LUX-1 year data)
increasing the sensitivity by a factor $5$. From 2018, the next
generation of experiments with multi-tonne Xenon (Xenon1T, LUX-Zeplin)
should reach cross-sections down to $10^{-47}$cm$^2$. {\it Indirect
  searches} aim at detecting the products of DM annihilation or decay
in the form of gamma-rays, neutrinos or antimatter cosmic rays. The
Fermi-LAT telescope, which is studying the gamma-ray flux from
multiple astrophysical sources, has provided the first strong
exclusion limits on DM annihilation cross-sections excluding
$\sigv\sim 3 \times10^{-26}$ cm$^3$/s for candidates with masses up to
$\sim 100$ GeV for a 100\% annihilation into $\bar b
b$~\cite{Ackermann:2013yva,Anderson:2014}. It is very likely that new
data will soon be available with e.g. AMS-02 for antimatter cosmic
rays, Ice-Cube for neutrinos, HESS-II for gamma-rays and with a new
generation of experiments, such as Gamma-400 and the CTA. From 2018
onwards, see e.g.~\cite{Bringmann:2012ez} for a review, the new
generation of indirect detection experiments will improve considerably
the current results. {\it Collider searches} can also constrain DM
scenarios through missing energy searches, although in most cases the
constraints are strongly model dependent. In addition, if dark matter
has non-negligible couplings to the Higgs field, the current
measurement of the Higgs decays can also be relevant for testing DM
scenarios~\cite{Zhou:2014dba,Aad:2014iia,Chatrchyan:2014tja}.
Finally, {\it astro-cosmo probes} can provide complementary
constraints, testing e.g. distortions of the Cosmic Microwave
Background (CMB) signal due to energy injection from DM annihilation
in the early universe, see
e.g.~\cite{Ade:2013zuv,Lopez-Honorez:2013cua,Madhavacheril:2013cna,Gallitalk}.

All the data provided by the complementary searches listed above
strongly constrains dark matter models. In particular, the composite
Higgs model $SO(6)/SO(5)$ can only satisfy all the current DM and LHC
constraints if the compositeness scale ($F$) is $\gtrsim 1$ TeV and if
$ m_{\text{DM}} \gtrsim 200$ GeV, as discussed in detail in
refs.~\cite{Frigerio:2012uc, UrbanoCDM:14}.  Due to the large amount
of data available from direct, indirect and LHC searches, it is
possible to formulate even more general statements on the possible
composite scenarios, if these are required to provide a DM candidate
behaving as a WIMP and accounting for all the DM abundance.  This is
the main purpose of this work. Here we consider the class of composite
models where the only composite states present in the low energy
effective theory are the Higgs doublet and the DM multiplet, and both
are pNGBs.  The possible strong resonances and remaining pNGBs are
assumed to be decoupled in the effective Lagrangian, either because
they are significantly heavier than the Higgs and the DM multiplet or
because they correspond to gauge degrees of freedom, due to a partly
gauged $\mG$.  We discuss this possibility with a minimal knowledge of
the UV completion.  In particular, we do not specify the fermion
representations under $\mG$, but parametrize their effects instead.
Within this framework, any composite Higgs model containing a DM
candidate can be mapped at low energies ($\ll \Lambda \simeq 4\pi F $)
to an effective Lagrangian containing the Standard Model (including
the Higgs doublet) and the DM multiplet.

In section~\ref{compDM} we define the class of models we will consider
and present the basic framework used to compute our results.  In
section \ref{sec:results} we classify the different scenarios
according to the DM representation under $SO(4)$ and discuss each case
separately. The simplest case, consisting of a singlet DM, is
discussed in detail in section~\ref{sec:singletRes}.  A minimal
realization of this scenario can be obtained for $\mG/\mH =
SO(6)/SO(5)$, as shown in refs.~\cite{Frigerio:2012uc, UrbanoCDM:14}.
The next simplest scenario consistent with $SO(4) \subset \mH $ is the
complex doublet DM case, discussed in section~\ref{sec:doubletRes}.
Although we once again present our results in a model independent way,
we also discuss a realization of the doublet DM case, corresponding to
the $SO(6)/ SO(4)\times SO(2)$ coset.  In section~\ref{sec:tripletRes}
we present the results for the triplet and general constraints on
higher representations. Finally, in section~\ref{conclusions} we
summarize our results and present the conclusions.  In the
appendix~\ref{appD6} we define our notation and discuss in detail the
derivation of the effective Lagrangians used in our calculations.

\section{General Composite Dark Matter}
\label{compDM}

In composite Higgs models \cite{TCHM:11, BeyondMCHM:09, Dim6Deriv:11, Alonso:2014, Carmona:2014iwa, Ryttov:2008xe, Alanne:2014kea}, the strong
interactions responsible for compositeness are assumed to break a
global symmetry group $\mG$ down to a smaller symmetry group $\mH$ at
the $F$ scale. The low energy spectrum, at energies $ \ll \Lambda
\simeq 4\pi F$, consists of the massless Nambu-Goldstone Boson (NGB)
modes with an effective Lagrangian completely fixed by the symmetry
breaking pattern $\mG \to \mH$.  For a general $\mG/\mH$ breaking, the
effective interactions of the Nambu-Goldstone modes to lowest order in
$\mathcal{O}(1/F)$ are given by the Callan, Coleman, Wess, Zumino
(CCWZ) Lagrangian \cite{CCWZ1:69, CCWZ2:69}. In order to avoid large
corrections to EW precision observables, we assume
$SO(4) \subset \mH $, which implies that the CCWZ Lagrangian
considered here is $SO(4)$ invariant.

If one of the NGBs must play the role of the Higgs doublet, $\mG$ must
be explicitly broken, resulting in a non-flat potential for the
pseudo-Goldstone fields. Here we assume the partial compositeness
scenario, where the explicit breaking is restricted to Yukawa and
gauge interactions between the SM and the strong sector. Once these
interactions are determined, the effective scalar potential for the
pNGBs can be computed up to form factors describing the strong
dynamics.  Since the effective scalar potential is generated only at
loop level, the pNGBs are expected to have masses $\lesssim F$.  The
fine-tuning in the Higgs potential is given by the separation between
the $F$ and EW scales. A fine-tuning not worse than 1\% implies:
\begin{equation}
F \lesssim 3 \mbox{ TeV}\,,
\end{equation}
but the upper bound in realistic models is usually
stronger~\cite{Panico:2012uw}.  Currently the strongest lower bound on
$F$ comes from EW precision tests and direct searches for the strong
resonances at the LHC, typically resulting in $F>800$
GeV~\cite{SeTFit2}. Although one may argue that incalculable UV
effects may hinder the impact of EW precision data on compositeness,
at this point Higgs data alone can set $F \gsim 700$
GeV~\cite{Falkowski:2013dza}.

In the MCHM \cite{MCHM:05}, where $\mG/\mH = SO(5)/SO(4)$, the four
pNGBs correspond to the degrees-of-freedom of the Higgs
doublet. Larger cosets can provide a stable DM candidate, as shown in
refs.~\cite{Frigerio:2012uc, UrbanoCDM:14, Chala:2012}.  In such
cases, in order for one of the pseudo-Goldstone bosons to be a viable
DM candidate, we must assume that both the strong sector and the SM
respect an exact (or approximate) $\mathbb{Z}_2$ symmetry, so the
lightest $\mathbb{Z}_2$-odd pNGB is (quasi-)stable.\footnote{Notice
  that the $\mathbb{Z}_2$ symmetry could already be present within the
  $\mG$ global symmetry group, see e.g.\cite{Frigerio:2012uc}.}
Furthermore, since current LHC data points to a SM-like Higgs, we
assume the Higgs to be the lightest $\mathbb{Z}_2$-even pNGB,
transforming as a bi-doublet under $SO(4)$. On the other hand, the DM
candidate can in general be a mixture of distinct $SO(4)$-multiplets
after electroweak symmetry breaking (EWSB). Here we only consider the
cases where the DM state (mostly) belongs to a single $SO(4)$
multiplet, so the mixing with other multiplets can be
ignored.\footnote{ In models where, after EWSB, there is a significant
  mixing between the lightest $\mathbb{Z}_2$-odd multiplets, the
  lightest $\mathbb{Z}_2$-odd state is usually charged, so there is no
  viable DM candidate, see e.g.~\cite{Hambye:2009pw} in the
  non-composite case.} Therefore, the low energy spectrum consists of
the Higgs doublet and the DM multiplet.  We assume all possible
remaining pNGBs and strong sector resonances to be decoupled in the
effective Lagrangian. This approximation is exact in minimal cosets,
where the Higgs and the DM multiplet correspond to all the pNGBs, as
in the particular examples discussed in section~\ref{sec:results}.
Additionally, note that higher derivative terms in the
  $\mG/\mH$ chiral Lagrangian, such as the Wess-Zumino-Witten term
  \cite{BeyondMCHM:09}, can potentially break the $\mathbb{Z}_2$
  symmetry.  Since these terms strongly depend on the coset $\mG/\mH$
  and the representation of the fermions under $\mG$, which we do not
  specify, they can not be computed in our model independent
  approach. Hence, we make the simplifying assumption that only those
  that preserve the $\mathbb{Z}_2$ symmetry are present, so the
  stability of DM is maintained. Furthermore, there are particular
  choices of cosets $\mG/\mH$ and fermion representations, where the
  $\mathbb{Z}_2$--violating interactions are zero and the $\xi
  \rightarrow-\xi$ parity is unbroken, see refs.
  \cite{Frigerio:2012uc, UrbanoCDM:14} for a specific realization.

\begin{figure}[h!]
  \begin{center}     
     \includegraphics[width=13cm]{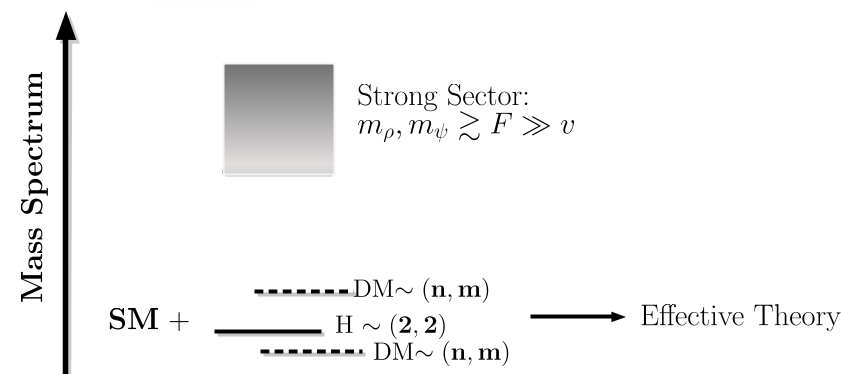}
  \end{center}
  \caption{Schematic representation of the spectrum for the class of models
  considered here. At low energies ($\ll F$) the effective theory we
  consider only contains the SM (including the composite Higgs bi-doublet) and
  the DM multiplet. The two possibilities for the DM mass are shown: $m_{DM} >
  m_h$ or $m_{DM} < m_h$.}
\label{fig:compDMsch}
\end{figure}

Any composite Higgs model satisfying these assumptions, detailed in
the appendix~\ref{appD6}, can be mapped to a low energy effective
theory valid for scales $\ll \Lambda$, as shown schematically by
figure~\ref{fig:compDMsch}.  Below we discuss the general form of the
effective operators relevant for computing the DM observables and
their impact on the DM phenomenology.

\subsection{Aspects of the Effective Dark Matter Lagrangian}
\label{effLagrangian}

The most general effective Lagrangian consistent with the assumptions
discussed in the previous section (see appendix~\ref{appD6} for more
details) contains a large number of unknown parameters. The latter
correspond to the coefficients of all the possible dimension six
operators involving SM fields as well as those involving the DM
multiplet. Fortunately, only a small subset of operators is relevant
for computing the DM observables.  We also consider the minimal
scenario, where DM is a cold thermal relic and the observed DM
abundance is generated through the standard freeze-out mechanism.  We
only consider tree level diagrams in our calculations. In addition,
processes where DM annihilates into more than two final states, such
as $\xi + \xi \to X + Y + Z$ are usually phase space suppressed with
respect to $\xi + \xi \rightarrow X + Y$.  Hence we neglect operators
with more than four fields. Notice though that in some cases,
annihilation into three body final states are well known to be
important~\cite{Chen:1998dp,Hosotani:2009jk,Yaguna:2010hn,Honorez:2010re,Belanger:2013oya}. In
particular, here we will take into account the contribution of
annihilations into off-shell gauge bosons $\xi + \xi \to V + V^*$.
Let us also emphasize that dimension five and six operators can become
relevant if they contain one or more Higgs fields, since after EWSB
these operators generate 4-field operators (with $H \to \langle H
\rangle = v$).  Furthermore, the $\mathbb{Z}_2$ symmetry, under which
$\xi$ is odd and $H$ is even, requires all operators to contain an
even number of $\xi$'s.

Given all the assumptions above, the operators containing more than
two powers of $\xi$ do not affect the relic abundance or the detection
rates. We just need to consider operators with zero or two DM
fields. Hence, for convenience, we split the effective DM Lagrangian
in a $\xi$-independent sector ($\mathcal{L}^{(0)}$) and a part
containing two powers of $\xi$ ($\mathcal{L}^{(2)}$): \be
\mathcal{L}_{\text{eff}} = \mathcal{L}^{(0)}(H,\ldots) +
\mathcal{L}^{(2)}(H,\xi,\ldots) , \ee with $\mathcal{L}^{(0)}$ given
by \be \mathcal{L}^{(0)}(H,\ldots) = \mathcal{L}_{\text{SM}} +
\mathcal{L}_6, \ee where $\mathcal{L}_{\text{SM}}$ is the SM
Lagrangian and $\mathcal{L}_6$ contains the relevant dimension-6
operators induced by the strong dynamics.  Since $\mathcal{L}_6$ is
independent of $\xi$, its operators are common for all the cases to be
discussed later.  The SM dimension-6 operators have been explored at
length in the literature \cite{SILH:07, EffHiggsCorbett:13,
  EffHiggsContino:13, EffHiggsElias:1302, EffHiggsElias:1308, Alonso:2012} and are
briefly discussed in appendix~\ref{appD6}. Below we simply list the
higher dimensional operators considered in our analyses. We use
\begin{equation}
  {\cal L}_6 =\frac{a_{2H}}{F^2} \left(\dmu |H|^2 \right)^2 -
  \frac{\lambda_1 \lambda_{H6}}{F^2} |H|^6  - \frac{c_4}{F^2}|H|^2 \left[\left(y_t
    \bar{Q}_L H^c t_R + y_b \bar{Q}_L H b_R  \right) + \text{h.c.}\right]\, ,
  \label{eq:l0}
\end{equation}
where $a_{2H}$, $c_4$ and $\lambda_{H6}$ are generated after we
integrate out the strong sector, while $\lambda_1$ is the dim-4
Higgs self-coupling.  As we assume a $CP$-even Higgs,
$c_{4}$ is real. The coefficients $a_{2H},c_4$ and $\lambda_{H6}$
  are  considered as ${\cal O}$(1) parameters. Notice that the operators 
in eq.~(\ref{eq:l0}) are weakly constrained by the LHC data 
\cite{EffHiggsCorbett:13, Ellis:2014jta}.

In order to reduce the number of parameters that would
  potentially affect the DM observables and simplify our analysis, we
  consider one single coefficient $c_4$ multiplying the dimension-6
  operators involving a direct coupling between the Higgs particle and
  the top and bottom quarks. As far as relic abundance and indirect
  dark matter detection searches are concerned, our choice could be
  justified by the fact that this extra coupling to top and bottom
  quarks are relevant in different regions of the parameter space. The
  impact of similar operators involving light quarks would be
  suppressed by the Yukawas ($y_\psi$). Let us emphasize though that
  this is however not true for direct dark matter searches in which
  case the relevant Higgs-nucleon coupling is dominated by heavy
  quarks ($b,t$ and $c$) contributions. The latter are due to
  heavy quark currents coupling to gluons through triangle diagrams
  involving heavy quark loops, see e.g.~\cite{Jungman:1995df} and
  references therein\footnote{Let us mention though that the
    contribution to the amplitude of DM-nucleon scattering related to
    the coupling $c_4$ is always expected to be a subdominant
    correction compared to other operators as it appears in the
    following combination: $\lambda_{\xi H}(1+c_4 v^2/F^2)$ (where
    $\lambda_{\xi H}$ is the Higgs-DM coupling). We expect thus the
    impact of choosing a common dimension-6 coefficient $c_4$ to all
    quarks to be marginal. }.
  
 As it is well known, after EWSB the operator proportional to $a_{2H}$
 generates a non-canonical kinetic term for the Higgs field. The
 rescaling to a physical Higgs introduces the factor
\begin{equation}
  R = 1/\sqrt{1 + 2 a_{2H} \frac{v^2}{F^2}} \, 
  \label{eq:Rfactor}
\end{equation}
in all Higgs interactions. Even though we include $R$ in our numerical
calculations, for simplicity we will typically neglect it in the analytic
expressions. Indeed $R$ is $ \simeq 1$ for $F \gtrsim 1$ TeV.

Notice that we do not consider the dimension-6 operators that modify
the gauge couplings to fermions, such as $( H^\dagger\sigma^j
{\overleftrightarrow D^\mu} H ) \left( \bar{Q}_L \sigma^j \gamma_\mu
Q_L \right)$, and those which contribute to the electroweak parameters
at tree level, like $\left( H^\dagger\sigma^j H \right)(W^j_{\mu
  \nu}B^{\mu \nu})$ \cite{SILH:07, EffHiggsCorbett:13,
  EffHiggsContino:13, EffHiggsElias:1302, EffHiggsElias:1308}.  As it
is well known, these operators are strongly constrained by the
electroweak fit~\cite{SeTFit2}.

Since we  assume that $\xi$ is a $SO(4) \sim SU(2)_L \times
SU(2)_R$ multiplet, the  possible representations for $\xi$ are
\be
(2 j_L + 1, 2 j_R + 1) = \left(1,1\right), \left(2,2\right), \left(3,1\right),
\left(1,3\right), \left(3,3\right), \left(2,4\right),\ldots\, .
\ee
In the following we limit our results to the cases
\begin{itemize}
  \item Singlet DM: $\xi \sim \left(1,1\right)$
  \item Doublet DM: $\xi \sim \left(2,2\right)$
  \item Real Multiplet DM ($\xi^C = \xi$): $\xi \sim \left(n,1\right)$, where
  $n=3, 5,\ldots$ .
\end{itemize}
Our restriction to real representations (except in the doublet case)
is justified by the fact that the phenomenology associated to complex
multiplets would be very similar to the real multiplet case; the
complex multiplet being equivalent to two degenerate real
multiplets. A real DM multiplet also implies that $\xi$ is not charged
under any $U(1)$ symmetry, since $\xi^C = C \xi^* = \xi$.
Consequently, $\xi$ is not charged under $U(1)_Y$ and since the
hypercharge operator is a linear combination of $T^3_R \subset
SU(2)_R$ and $U(1)_X$\footnote{As it is well known in composite Higgs
  models, it is often necessary to enlarge the $\mG$ group with an
  additional $U(1)_X$ symmetry in order to obtain the correct
  hypercharges for the SM fermions~\cite{Mrazek:2011iu,SILH:07}.}, it
means that $\xi$ is a singlet under $SU(2)_R$.  Hence $\xi \sim
\left(n,1\right)$.

Finally we briefly comment on the validity region of the effective
Lagrangian used in our analysis. The energy involved in the relevant
DM processes is typically $\sqrt{s} \lesssim 2 m_{DM}$, where the
upper limit corresponds to non-relativistic annihilation at zero
velocity today. In the early universe, energies involved are usually
slightly larger given that $v\sim 0.2$ while for direct searches the
energies involved are much smaller. Therefore, we expect the effective
Lagrangian approach used here to be valid as long as the new physics
states (which could affect the DM observables) are well above $2
m_{DM}$. In composite Higgs models one naively expects the strong
resonances to have masses of order $\Lambda \simeq 4 \pi F$. However,
the composite partners of the top quark must be lighter in order to
avoid re-introducing a large fine-tuning in the Higgs
potential~\cite{Panico:2012uw}.  Here we assume that the spinorial
resonances have masses $ m_{\Psi} \gtrsim F$, hence our effective
Lagrangian is valid for $m_{DM} \ll m_{\Psi}$ or $m_{DM} \ll F$, which
we use as a strict upper limit in our study.  It is also important to
point out that, for $m_{DM} < F$, all the operators used in our
analysis are far away from any unitarity
bounds~\cite{Griest:1989wd,Corbett:2014ora}.

\subsection{The Effect of Higher Dimensional Operators}\label{effects}

The specific form of the dimension-4  and 6 relevant
operators\footnote{We always mean by dimension-6 operators, the
  operators suppressed by $1/F^2$.} consistent with $ SO(4) \subset
\mH $ are derived in detail in the appendix~\ref{appD6} and will be
discussed in section~\ref{sec:results}. It is instructive, however, to
first comment on the general properties of these operators and how
they can affect the DM observables.  Since we are only interested in
4-field operators after EWSB, the relevant dimension-6 operators
generated by integrating out the strong sector heavy resonances are of
the following types
\begin{align}
\frac{y_f}{v}\frac{v^2}{F^2}\xi^2 \bar{\psi} \psi,\;
\lambda \frac{v^3}{F^2}\xi^2 h,\; \lambda \frac{v^2}{F^2}\xi^2 h^2,\; g^2
\frac{v^2}{F^2}\xi^2 V_\mu V^\mu,\;g \frac{v p_\mu}{F^2}V_\mu h,\; \frac{v
p^2}{F^2} \xi^2 h \mbox{ and }
\frac{p^2}{F^2} \xi^2 h^2, \label{genops}
\end{align}
where $h$ is the Higgs boson field, $V_\mu$ represents a
$SU(2)_L\times U(1)_Y$ gauge field, $\psi$ is a SM fermion, $g$ is the
corresponding gauge coupling, $\lambda$ represents the corresponding
dimension-4 coupling and $p_\mu$ refers to the typical momentum scale
for the process.  As all the above operators are presumed to have
coefficients $\lesssim 1$, one would expect such operators to
be irrelevant with respect to the corresponding dimension-4 operators
for $F \gg v$.  However it is often the case that, in order to
generate the correct relic abundance, the dimension-4 operators
coefficients need to be suppressed. In such cases, the dimension-6
operators may still play a significant role. Furthermore, the
derivative operators can be enhanced if $p^2 \sim m_{DM}^2 \gg
v^2$.

\begin{figure}[h!]
  \begin{center}     
     \includegraphics[width=7.5cm]{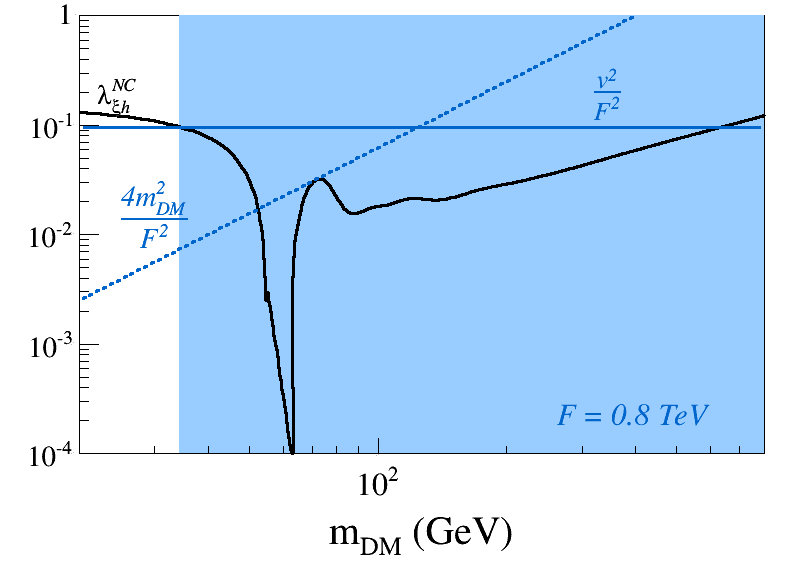}
     \includegraphics[width=7.5cm]{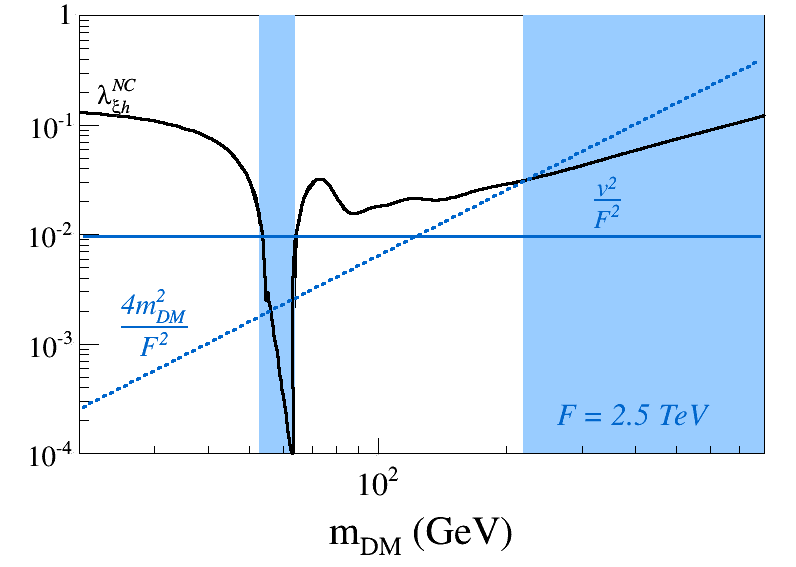}
  \end{center}
  \caption{{\it Singlet DM model:} The black continuous line denotes
    the values of the DM-Higgs coupling  in the
    non-composite case ($\lambda_{\xi h}^{NC}$) as a function of $m_{DM}$ giving
    rise to the right relic abundance. The blue shaded areas correspond to the windows of the parameter space in which
    the dimension-6 operators of eq.~(\ref{genops}) can potentially
    affect the DM observables for $F = 0.8$ TeV (left) and $F = 2.5$ TeV
    (right).  In the regions below the continuous
    horizontal blue line $\lambda_{\xi h}^{NC}\lesssim v^2/F^2$ while in the
    regions below the blue dotted curves $\lambda_{\xi h}^{NC}\lesssim 4
    m_{DM}^2/F^2$.}
\label{fig:ops}
\end{figure}

In order to illustrate the above arguments more concretely, we take
the singlet DM case as an example.  In the non-composite limit ($F \to
\infty$), there is only one relevant dimension-4 operator,
corresponding to the Higgs-DM coupling ($\lambda_{\xi h}$), also known
as the Higgs portal coupling. As a result, the parameter space in the
non-composite singlet case consists only of $\lambda_{\xi h}$ and
$m_{DM}$.  As it is well known, requiring the DM relic energy density
$\Omega_{DM} h^2 \simeq 0.12$~\cite{Hinshaw:2012aka,Ade:2013zuv} fixes
the Higgs-DM coupling for a given DM mass and the solutions with the
correct relic abundance correspond to a line in the $\lambda_{\xi
  h}$-$m_{DM}$ plane.  This is shown by the black solid line in
figure~\ref{fig:ops}, where we plot the value of the Higgs-DM coupling
in the non-composite case ($\lambda_{\xi h}^{NC}$) required to
generate the correct relic abundance.  As we can see, $\lambda_{\xi
  h}^{NC}$ does indeed take values $\ll 1$ in most of the parameter
space, which may render the operators in eq.~(\ref{genops}) relevant
in these regions even though they are suppressed by $1/F^2$.  In order
to estimate for which values of $m_{\text{DM}}$ the inclusion of the
dimension-6 operators may affect the DM observables, we compare their
potential effective strength with respect to $\lambda_{\xi h}^{NC}$.

Assuming that the coefficients multiplying the operators in
eq.~(\ref{genops}) are $\mathcal{O}(1)$, the blue shaded areas in
figure~\ref{fig:ops}, for $F = 0.8$ TeV (left) and $F = 2.5$ TeV
(right), correspond to the windows of the parameter space in which
these operators can potentially drive the relic abundance.  In the
regions below the continuous horizontal blue line $\lambda_{\xi
  h}^{NC}\lesssim v^2/F^2$, i.e. the non-derivative operators can be
dominant, while the regions below the blue dotted curves $\lambda_{\xi
  h}^{NC}\lesssim 4 m_{DM}^2/F^2$, i.e. the derivative operators can
become dominant. As we can see, for $F = 0.8$ TeV, the dimension-6
operators can {\it potentially} affect the DM cross-sections for
$m_{DM} \gtrsim 35$ GeV.  On the other hand, for higher values of $F$
this contribution becomes limited to the Higgs resonance, where
$\lambda_{\xi h}^{NC} \ll 1$, or the heavy DM region, where the
derivative operators can be enhanced.  Furthermore, for $F = 2.5$ TeV,
we see that the non-derivative operators play no role, except around
the resonance.  We stress that the results shown in
figure~\ref{fig:ops} are purely schematic and are intended only to
illustrate the {\it potential} impact of including the higher
dimensional operators in our analyses. However, as it will be shown in
the next section, the overall features described above do appear in our numerical computations.

It is also interesting to point out that the dimension-6 operators can
have very distinct contributions to the annihilation and nucleon
scattering rates.  On the one hand, while the momentum
  transferred in DM-nucleon scattering is $\mathcal{O}$(keV-MeV), the
  momentum scale involved in DM annihilations in the early universe is
  of order $m_{\text{DM}}$. Hence, the derivative operators which
  scale as $p^2/F^2$ only contribute significantly to DM annihilations
  and not to the direct detection rate. On the other hand, dimension-6
  operators involving Yukawa type of interactions such as
  \begin{equation}
    \frac{C_q}{F^2}|\xi|^2 \left(y_q \bar{Q}_L H^c q_R \right)\,,
    \label{eq:Hyuk}
  \end{equation}
  where $y_q$ is the Yukawa coupling of the $q$-quark and $C_q$ a
  coefficient of $\mathcal{O}(1)$, give important contributions to the
  dark matter-nucleon scattering cross section, $\sigma_{DMp}$, for
  $q=c,b$ and $t$ (through triangle diagrams processes that are, by
  default, taken into account using dark matter tools such as {\tt
    micrOMEGAS}, see~\cite{Belanger:2008sj}).

Let us mention that, in the following, we have considered interactions
of the type of (\ref{eq:Hyuk}) with couplings to the 3rd generation of
quarks only ($C_{u,d,c,s} = 0 $). This assumption does not affect
  the relic density and the indirect detection results, since the
  contributions from light quarks is always suppressed by their
  Yukawas.  The couplings to light quarks can however impact
  $\sigma_{DMp}$, especially in the regions where $\lambda_{\xi h}$ is
  small.  Given the values of the parameters driving the nucleon
  matrix element considered in our analysis\footnote{We took the
    default values of {\tt micrOMEGAS3.3} corresponding to $f_d^p=
    0.0191, f_u^p= 0.0153$ and $f_s^p= 0.0447$ in the notations of
    ref.~\cite{Belanger:2008sj}.}, we expect that including a coupling
  to all quarks would at most enhance $\sigma_{DMp}$ by a factor of
  $\sim 4$ and only if $|\lambda_{\xi h}|\ll m_h^2/F^2 |C_q| $ (see
  also~\cite{UrbanoCDM:14} for a similar discussion).  In order
    to reduce the number of free parameters in our analysis, we have
    however neglected this possibility.  Finally, we make one
  additional simplifying assumption: $C_t = C_b$.  Once again, this
  assumption does not affect the relic density or direct detection
  rate, since annihilations to tops or bottoms are relevant at
  distinct regions of parameter space. While $\sigma_{DMp}$ could
  change for non-universal couplings, we do not expect that our
  results are significantly affected by this assumption, since we
  expect $C_t \sim C_b \lesssim \mathcal{O}(1)$.

\section{Results}
\label{sec:results}

Using the effective Lagrangians derived in appendix~\ref{appD6}, we
now compute the DM observables and discuss how this constrains the
effective operators for a given DM representation.  The DM relic
abundance and other observables have been computed using
FeynRules~\cite{Alloul:2013bka} and {\tt
  micrOMEGAS}~\cite{Belanger:2014vza}. The calculation of the relic
abundance includes all possible annihilation channels into two body as
well as annihilation into off-shell $V= W$ or $Z$ boson ($\xi + \xi
\to V^* + V$), which can be relevant for $m_{DM} \lesssim m_V$.
Even though radiative corrections are already known to affect DM
  observables in similar non-composite DM scenarios, especially the
  dark matter-nucleon scattering cross
  section~\cite{Cirelli:2005uq,Klasen:2013btp,Abe:2015rja}, it is
  beyond the scope of this paper to evaluate all loop contributions
  within the composite framework (the exception being the well known
  contributions from triangle diagrams of heavy quark loops included
  by default in {\tt micrOMEGAS} for the computation of the DM-nucleon
  scattering cross section).  One should keep in mind though that
such corrections are expected to give rise to the lower bound
$\sigma_{DM\,p}\gsim 10^{-47} $ cm$^2$ for non-composite models in
e.g. the doublet case~\cite{Klasen:2013btp}.

In order to be as model independent as possible, we parametrize the
effective Lagrangian according to the results in appendix~\ref{appD6}
and scan over the ${\cal O}(1)$ coefficients of the dimension-6
operators in the range $\left[-1,1\right]$.  The unknown
  parameters for effective dimension-4 operators (denoted with
  $\lambda$ couplings) in the DM Lagrangian are also allowed to vary
  freely within a range restricted to $\left[-4 \pi,4 \pi\right]$.
We also require all the solutions to satisfy:
$0.0941<\Omega_{DM}h^2<0.127.$~\footnote{ We used the 2$\sigma$ WMAP
  5-year range~\cite{Hinshaw:2008kr}. For a reference, the 3$\sigma$
  WMAP 9-year range is
  $0.1003<\Omega_{DM}h^2<0.1273$~\cite{Hinshaw:2012aka} and the
  3$\sigma$ PLANCK range is
  $0.1103<\Omega_{DM}h^2<0.1289$~\cite{Ade:2013zuv}.}

In addition to the above conditions, directly taken into account in
the scans, we will superimpose several constraints from LHC, direct
and indirect DM searches. Furthermore, the low DM mass region is
limited by the Higgs invisible width, which has been recently
constrained by ATLAS \cite{Aad:2014iia} and CMS
\cite{Chatrchyan:2014tja}. Here we use the 95\% C.L. upper limit from
CMS: BR$(h \to \rm{invisible}) < 0.58$.
For the direct detection constraints, we consider the latest LUX
results \cite{Akerib:2013tjd} on the spin-independent DM-nucleon
elastic cross-section as well as the projected sensitivity of Xenon1T,
both at 95\% CL. The DM annihilation into SM particles in our galaxy could
copiously produce stable particles, such as photons, neutrinos, positrons or
  antiprotons.  In that framework, indirect detection searches
  constraints from Fermi-LAT~\cite{Ackermann:2013yva,Anderson:2014}
  and PAMELA~\cite{Adriani:2011cu,Adriani:2012paa} can be relevant
  here.  Indeed, a previous analysis of composite DM
  models~\cite{UrbanoCDM:14} showed that the bound on the DM
  annihilation cross-section derived from PAMELA measurements of the
  antiproton spectrum could provide constraints complementary to LHC
  and direct detection searches. The bounds associated to charged
  cosmic rays are, however, well known to suffer from astrophysical
  uncertainties~\cite{Evoli:2011id,Cirelli:2014lwa} affecting the
  propagation model.  Here we use the $3\sigma$ upper bound
  constraints derived in ref.~\cite{Cirelli:2014lwa} (figure 6c) for
  the DM annihilation cross-section into $\bar{b} b$.~\footnote{The
    PAMELA constraints are thus imposed on $\sigma v
    (\xi\xi\rightarrow \bar b b)|_{v=0}$ rescaling the constraints of
    ref.~\cite{Cirelli:2014lwa} by the associated branching ratio.
  } Let us mention that such a bound has been derived assuming an
  uncertainty of 50\% in the solar modulation of the antiproton flux
  and we take the limits associated to the so called CON propagation
  model (green curve in figure 6c of ref.~\cite{Cirelli:2014lwa}),
  which usually corresponds to conservative bounds.  Let us emphasize
  that even though similar bounds were derived for other annihilation
  channels such as e.g. $\xi \xi\to W^+ W^-$, we checked that they
  provide weaker constraints than the direct detection searches.
  We have thus neglected them.  Let us also mention that the current
  preliminary constraints on gamma-ray flux from the
  Fermi-LAT~\cite{Anderson:2014} measurements provide even stronger
  bounds on $\sigma v (\xi\xi\rightarrow \bar b b)|_{v=0}$. We have checked 
  that the impact of these preliminary results are negligible, so we only
  consider the constraints from PAMELA.

  The LHC constraints from monojet and monophoton searches are still
  insufficient to compete with the direct detection or invisible Higgs
  decay searches.  Nonetheless, for both the doublet and triplet
  representations, the DM multiplet contains charged states, which can
  decay to DM and be detectable at the LHC. The main final states are
  then missing $E_T$ plus $W$'s, $Z$'s and/or Higgses. Using
  SModelS~\cite{Kraml:2013mwa} we were able to recast the LHC
  constraints for SUSY searches and apply them to the scenarios
  considered. We have explicitly computed the relevant production
  cross-sections and verified that the LHC constraints do not lead to
  any new excluded region. Therefore we will not consider them in the
  following results.

\subsection{Singlet DM} \label{sec:singletRes}
\label{sec:singlet-dm}
\subsubsection{The Generic composite singlet DM} 
\label{sec:gener-comp-singl}

The singlet DM case corresponds to the 
following effective Lagrangian (for details see
appendix~\ref{singletLag})
\begin{align}
\mathcal{L}^{(2)} & = \frac{1}{2} \dmu \xi \dMu \xi
 - \frac{1}{2}\mu_{\xi}^2 \xi^2 - \frac{\lambda_3}{2}\left(1 +
 \frac{\lambda_3'}{F^2} |H|^2\right) \xi^2 |H|^2  + \frac{a_{d1}}{F^2} \dmu
 \xi^2 \dMu |H|^2  \label{singLeff} \\
 & - \frac{1}{2}\left[ \frac{d_4}{F^2}\xi^2 \left(y_t \bar{Q}_L H^c t_R + y_b
 \bar{Q}_L H b_R \right) + \text{h.c.}\right]\, ,\nonumber 
\end{align}
resulting in the following DM mass after EWSB:
\begin{eqnarray}
   \mdm^2 &=& \mu_\xi^2 + \lambda_3 \left(1+
   \frac{\lambda_3'}{2}\frac{v^2}{F^2}\right) \frac{v^2}{2}\, .
 \label{eq:sgen-m}
\end{eqnarray}
The coefficients $a_{d1},d_4,\lambda_3'$ are taken to be real ${\cal O}$(1)
parameters, while the $\lambda_3$ coupling is allowed to 
vary in the window $[-4\pi, 4\pi]$ and $\mu_{\xi}$ is the DM bare
mass. Except for $a_{d1}$, which is fixed by $\mG/\mH$, all the other
coefficients depend on the fermion embedding in $\mG$. Therefore in
our model independent approach we take these to be free parameters.
As discussed in Sec.~\ref{effects}, we neglect dimension-6 operators involving
light quarks and we assume a universal coefficient ($d_4$) for the top and
bottom couplings.

For $m_{DM}< m_t$, the DM observables are mostly determined by the
effective $\xi(p_1)-\xi(p_2)-h(p_h)$ coupling, which in the composite
case is given by:
\begin{eqnarray}
   &\lambda_{\xi h}& =  \left(\frac{\bar{\lambda}}{2}  - a_{d1}
 \frac{p_h^2}{F^2}\right)\, ,  \label{l3effs}
\end{eqnarray}
where
\begin{eqnarray}
&\bar{\lambda} = \lambda_3 \left(1 + \lambda_3' \frac{v^2}{F^2}\right)
\label{lbar}
\end{eqnarray}
and $p_h^2$ is the (off-shell) Higgs momentum.

It is important to point out that the effective DM-Higgs coupling
relevant for the direct detection (DD) rate and the relic density (RD)
(or $\sigv$) can take very different values, due to the momentum
dependence in eq.~(\ref{l3effs}).  While for annihilations through the
Higgs $s$-channel we have $p_h^2 \simeq 4 m_{DM}^2$, for direct
detection we have $t$-channel scattering, hence $p_h^2 \simeq
0$. Therefore it is convenient to define the effective couplings in
each of these regimes:
\begin{align}
\lambda_{\xi h}|_{p_h \sim 2 m_{DM}} & \equiv \lambda_{\xi h}^{RD} =
\frac{\bar{\lambda}}{2} - 4 a_{d1} \frac{m_{DM}^2}{F^2} \, ,\nonumber \\
\lambda_{\xi h}|_{p_h \sim 0} & \equiv \lambda_{\xi h}^{DD} =
\frac{\bar{\lambda}}{2} \, , 
\end{align}
where $\lambda_{\xi h}^{RD}$ ($\lambda_{\xi h}^{DD}$) is the relevant
coupling for the calculation of the relic abundance (DD
cross-section\footnote{Let us mention that, since we have not
  considered DM contact interactions with light quarks, the $d_4$
  coupling has only a minor impact on the general DD picture. If
  couplings to light quarks were included, the results could be
  modified in some small regions of parameter space, where
  $|\lambda_{\xi H} |\ll |d_4| m_h^2/F^2$, see Sec.~\ref{effects} for
  a brief discussion.}).

In the low mass region, where annihilations to tops are closed, both
the DM-nucleon spin independent scattering cross-section,
$\sigma_{DMp}$, and the thermally averaged annihilation cross-section
times the relative velocity at the freeze-out time, $\sigv$, are
proportional to the Higgs-DM effective coupling:
\begin{align}
\sigv & \propto \left(\lambda_{\xi h}^{RD}\right)^2 \simeq 
\left(\frac{\bar{\lambda}}{2} - 4 a_{d1} \frac{m_{DM}^2}{F^2}\right)^2 \, , 
\nonumber
\\
\sigma_{DMp} & \propto \left(\lambda_{\xi h}^{DD}\right)^2 \simeq  
\left(\frac{\bar{\lambda}}{2}\right)^2 \, .
\label{sigsapp}
\end{align}
In analogy to the non-composite case, once we impose $\Omega_{DM} h^2
\simeq 0.12$, $\lambda_{\xi h}^{RD}$ is fixed for each value of
$m_{DM}$. Denoting by $\lambda_{\xi h}^{NC}$ the value of the coupling
required to generate the correct relic abundance in the non-composite
case (shown by the black curve in figure~\ref{fig:ops}), we have:
\begin{align}
\left(\lambda_{\xi h}^{RD}\right)^2 & \simeq \left(\lambda_{\xi h}^{NC}\right)^2
\To \left(\frac{\bar{\lambda}}{2}\right)^2 \simeq \left(\lambda_{\xi h}^{NC} \pm 4 
a_{d1} \frac{m_{DM}^2}{F^2} \right)^2\nonumber \\
\To & \sigma_{DMp} \propto \left(\lambda_{\xi h}^{NC} \pm 4  a_{d1}
\frac{m_{DM}^2}{F^2} \right)^2 \;\mbox{ or }\; 
\sigma_{DMp}/\sigma_{DMp}^{NC} = \left(1 \pm 4  \frac{a_{d1}}{\lambda_{\xi
h}^{NC}} \frac{m_{DM}^2}{F^2} \right)^2 \, ,
\label{sDDapp}
\end{align}
where $\sigma_{DMp}^{NC}$ is the DM-nucleus cross-section in the
non-composite case. From the above result we see that $\sigma_{DMp}$
can be either enhanced or suppressed with respect to its non-composite
value, depending on the values of $a_{d1}$ and $m_{DM}$. Let us
  stress that the above results are only approximate, since we have
  neglected the contributions from $d_4$, both for relic abundance and
  direct detection considerations. The impact of $d_4$ is,
    however, included in our numerical analysis.

\subsubsection{ Scan results}
\label{sec:scan-results-s}

We now present the full numerical results for the DM observables for a
random scan over the coefficients of the effective Lagrangian.  Since
in some cases the correct relic abundance requires extremely small
values of $\lambda_3,\lambda_3'$ and $\lambda_{H6}$, we scan
logarithmically over their allowed range. All the other parameters are scanned linearly within
the ranges:
\begin{eqnarray}   
  &10\mbox{ GeV}<\mdm<F \, ,\nonumber\\
  &-4\pi <\lambda_3<4\pi \, ,\nonumber\\
  &-1 <\lambda_3'<1 \, ,\nonumber\\
  &10^{-6} <\lambda_{H6}< 1 \, ,\nonumber\\
  &-1 <c_4, a_{2H}, d_4, a_{d1}<1 \, .
\label{eq:parm-s}
\end{eqnarray}

\begin{figure}[h!]
  \begin{center}     
     \includegraphics[width=7.5cm]{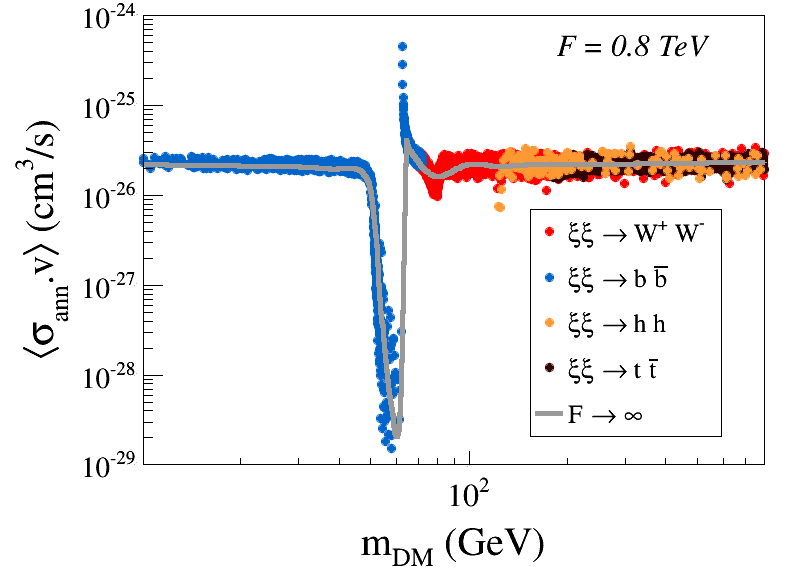}
     \includegraphics[width=7.5cm]{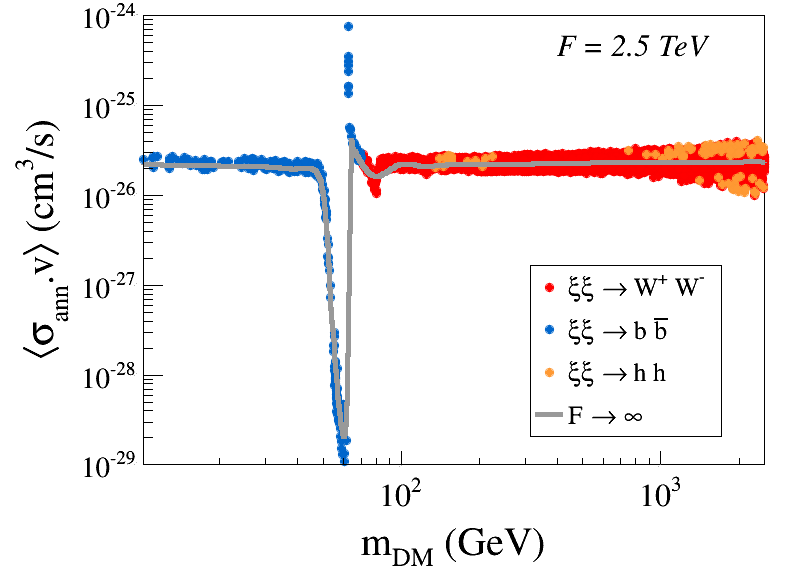}
  \end{center}
  \caption{{\it Generic composite singlet DM:} Values of the DM
      annihilation cross-section at zero velocity for $F = 0.8$ TeV
      (left) and $F = 2.5$ TeV (right) satisfying to $0.941 <
      \Omega_{DM} h^2 < 0.127$.  The channels giving rise to the
      largest branching ratio to the annihilation cross-section are
      shown with different colors. We also show as a solid gray line
    the value of $\sigv$ for the non-composite case ($F \to \infty$).}
\label{fig:sigv}
\end{figure}

First, in figure~\ref{fig:sigv}, we show the results for the DM
annihilation cross-section ($\sigv$) today at zero velocity 
as a function of $m_{DM}$. For comparison purposes we also show the
respective values for $\sigv$ in the non-composite case (solid gray
line). As we can see, for both $F = 0.8$ TeV (left plot) and 2.5 TeV
(right plot), the annihilation rate follows closely the non-composite
case, as expected from the relation $\Omega_{DM} h^2 \propto 1/\sigv$.
We notice, however, that while in the $F = 2.5$ TeV case the dominant
annihilation channels are the same as in the non-composite case ($\xi
\xi \to \bar{b} b$ or $\xi \xi \to W^- W^+$), for $F = 0.8$ TeV, both
annihilations to $\bar{t} t$ and $hh$ can be dominant at large DM
masses.  This is mostly due to the fact that, for small values of $F$,
the non-derivative operators $d_4 \xi^2 \bar{Q}_L H^c t_R$ can be
comparable to $\lambda_{\xi h}$, as schematically shown in the left
hand (LH) side plot of figure~\ref{fig:ops}. On the other hand, for $F
= 2.5$ TeV, these operators are always suppressed.

\begin{figure}[h!]
  \begin{center}
     \includegraphics[width=7.5cm]{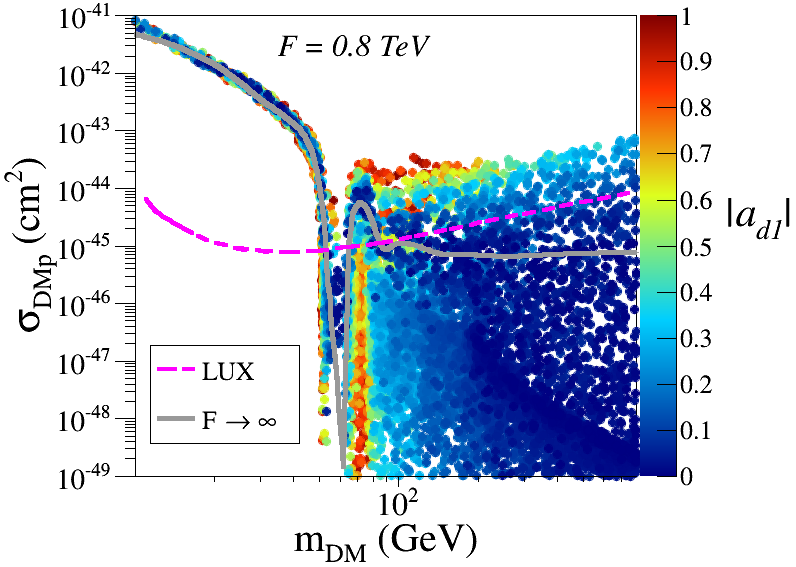}
     \includegraphics[width=7.5cm]{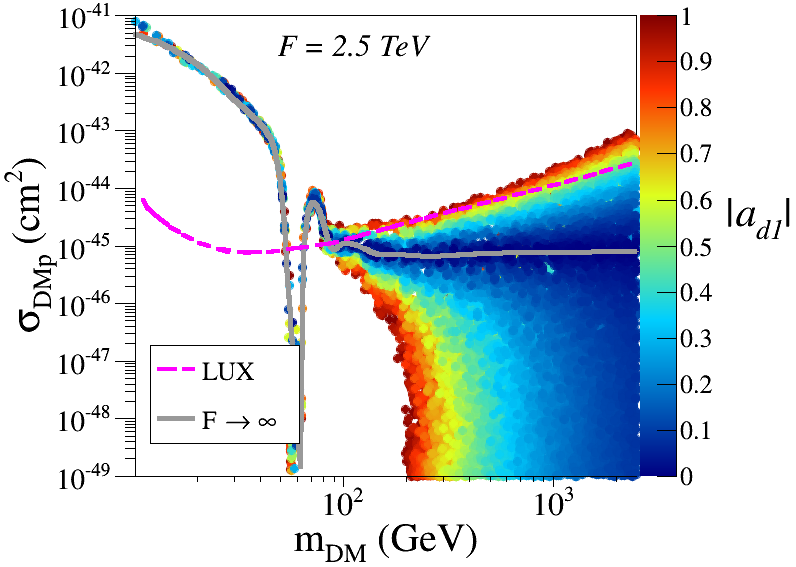}
  \end{center}
  \caption{ {\it Generic composite singlet DM:} Values of the
      DM-nucleon scattering cross-section as a function of
      $m_{DM}$. All points satisfy $0.0941 < \Omega_{DM} h^2 <
    0.127$. The absolute values of the dimension-6 coefficient
    $a_{d1}$ are shown with the gradient of colors. The LUX
    95\% CL bound is also shown.  }
\label{fig:sDD}
\end{figure}

The DM-nucleon scattering cross-section ($\sigma_{DMp}$) is shown as a
function of $m_{DM}$ in figure~\ref{fig:sDD} for $F= 0.8$ TeV (left)
and 2.5 TeV (right). All points give the correct DM relic abundance.
Once again the non-composite case is shown by the solid gray
line. Going from low to high $m_{DM}$ it is possible to
  recognize in figure~\ref{fig:sDD}
  some known features~\cite{Frigerio:2012uc,UrbanoCDM:14}: the Higgs resonance
  region ($m_{DM}= m_h/2$), the $\xi \xi \to WW$ threshold
  ($m_{DM}\simeq m_W$) and the high mass region ($m_{DM}\gsim m_t$).  We
also show by distinct colors the absolute value of the coefficient
$a_{d1}$ for each point.
 As we can see, the low mass region ($m_{DM} \lesssim 50$ GeV) is
 unaffected by the dimension-6 operators induced by the strong
 sector. This is already expected from the discussion in
 section~\ref{effects}, where we have shown that the low mass region
 is mostly insensitive to these operators.

Once $m_{DM} \gtrsim 50$ GeV, $\sigma_{DMp}$ can take values
drastically different from the non-composite case. As expected from
eq.~(\ref{sDDapp}), this is due to the contribution of the derivative
operator proportional to $a_{d1}$ (at least for $m_{DM} \lesssim
m_t$). Indeed, given that we can get the right abundance with $ 4
a_{d1} m_{DM}^2/F^2\simeq \lambda_{\xi h}^{NC} $, i.e. for small
  values of $\bar{\lambda}$, $\sigma_{DMp}$ can become very
suppressed, as seen in figure~\ref{fig:sDD}. In addition, since
$a_{d1}$ can take both negative and positive values, we can also have
the correct DM abundance with $\bar{\lambda}$ larger than in the
non-composite case, resulting in an {\it enhancement} of
$\sigma_{DMp}$, as shown by the points above the gray line in
figure~\ref{fig:sDD}.

For $m_{DM} > m_t$, annihilations to top pairs become kinematically
allowed and $\sigv$ receives a contribution from the operator
proportional to $d_4$.  In this case, we can once again generate the
correct relic abundance, even when $\lambda_{\xi h} \ll \lambda_{\xi
  h}^{NC}$.  This corresponds to all the points at large masses with
small values of $\sigma_{DMp}$ and $a_{d1}$ in the LH side plot of
figure~\ref{fig:sDD}.  Notice though that this feature is not as
strongly present in the right hand (RH) side plot of
figure~\ref{fig:sDD} for $F=2.5$ TeV. This is simply due to the fact
that for such large $F$ values and $m_{DM} > m_t$, non-derivative
operators, such as the one proportional to $d_4$, are always
subdominant, as already anticipated by the RH side plot of
figure~\ref{fig:ops}.  Finally, we also emphasize that for $F=2.5$
TeV, $\sigma_{DMp}$ only deviates significantly from its non-composite
value for higher DM masses ($m_{DM} \gtrsim 150$ GeV), as expected
from eq.~(\ref{sDDapp}) and in agreement with the discussion in
section~\ref{effects}.  From figure~\ref{fig:sDD} we already see that
for $m_{DM}> 200$ GeV to be consistent with LUX we need $\vert
a_{d1}\vert \lsim 0.2$ for $F = 0.8$ TeV, higher values are however
possible if we increase the scale of compositeness.

\subsubsection{Experimental constraints}
\label{sec:exper-constr-s}

We now discuss the current and future constraints on the generic
composite singlet DM scenarios.  As seen in figure~\ref{fig:sDD}, the
direct detection cross-section can be significantly suppressed or
enhanced in the composite case.  Nonetheless, the low mass region
still behaves mostly as Higgs portal DM, being strongly constrained by
direct, indirect DM searches and invisible Higgs decays.

\begin{figure}[h!]
  \begin{center}
     \includegraphics[width=7.5cm]{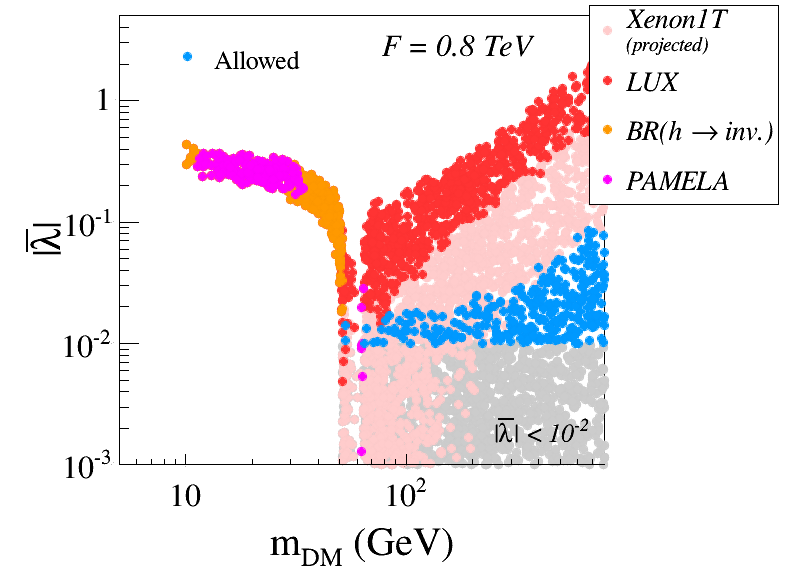}
     \includegraphics[width=7.5cm]{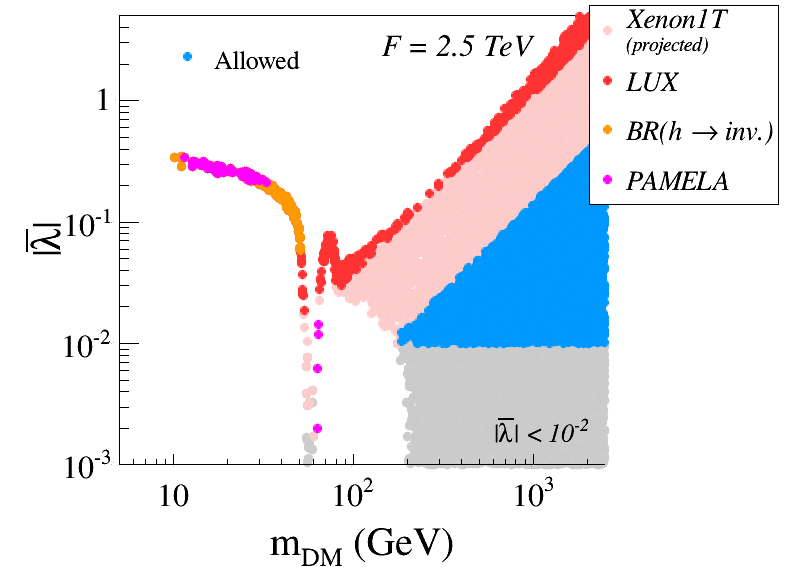}\\
     \includegraphics[width=7.5cm]{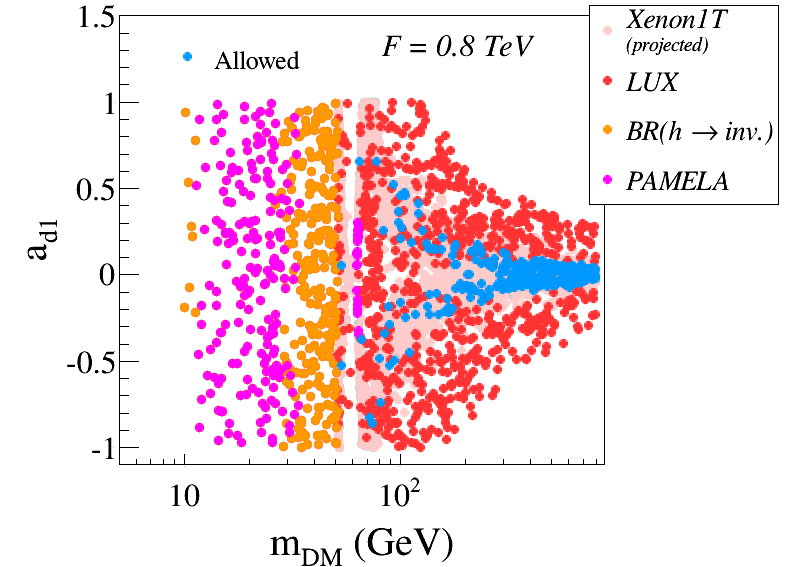}
     \includegraphics[width=7.5cm]{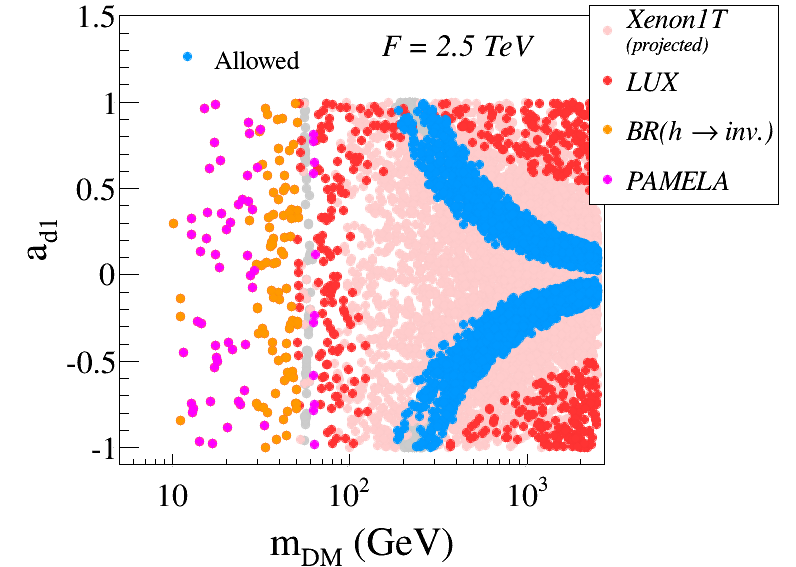}
 \end{center}
  \caption{ {\it Generic composite singlet DM:} Experimental
    constraints in the $|\bar{\lambda}|-m_{DM}$ (top) and
    $a_{d1}-m_{DM}$ (bottom) planes. We show results for $F = 0.8$ TeV
    (left) and 2.5 TeV (right).  See text for
    details.} \label{fig:constSinglet}
\end{figure}

We summarize the constraining potential of each of the
  experiments mentioned at the beginning of section~\ref{sec:results}
   in the planes $\bar{\lambda}-m_{DM}$ and $a_{d1}-m_{DM}$ of
  figure~\ref{fig:constSinglet}. Let us first mention that, in some
  regions of the parameter space, very small values of $\bar{\lambda}$
  are necessary to account for the right relic abundance.
Although this is typically consistent with most experimental
constraints, in the composite framework, both $\lambda_1$ (the
Higgs self-coupling) and $\bar{\lambda}$ are expected to be
generated by the explicit breaking of $\mG$ and to be of similar size, i.e.
$\mathcal{O}(\lambda_1) \sim \mathcal{O}(|\bar{\lambda}|) \simeq 0.1$.
An explicit example can be found in Ref.~\cite{UrbanoCDM:14}, where
$\bar{\lambda} \simeq 0.13$ for $\mG/\mH =
  SO(6)/SO(5)$. Therefore we consider the region with
$|\bar{\lambda}| < 10^{-2}$ as theoretically disfavored.

In figure~\ref{fig:constSinglet}, the top (bottom) plots illustrate,
in the plane $|\bar{\lambda}|-m_{DM}$ ($a_{d1}-m_{DM}$), for $F = 0.8$
(left) and 2.5 TeV (right), which regions are currently excluded by
direct DM searches (with LUX in red), by indirect DM searches (with
PAMELA in magenta) and due to large contributions to the Higgs
invisible width (in yellow).  The pink regions are currently allowed
and will eventually be probed by Xenon1T. The gray regions are
experimentally allowed by current data and outside the planned
sensitivity of Xenon1T, but they are theoretically disfavored. In the
bottom plots (for $a_{d_1}$) most of the gray points are hidden 
behind the region that will be probed by Xenon1T.

As expected from the discussion above, the low mass region ($m_{DM}
\lesssim 100$ GeV) is almost entirely excluded either by indirect,
direct searches or the Higgs invisible width constraints. Let us
emphasize that this is valid for both small and large values of
$F$. The only points still allowed in the low mass region lie around
the Higgs resonance and will be tested by Xenon1T.  However most of
these points have $\vert \bar{\lambda}\vert \lesssim 0.01$, which is
difficult to realize in specific models without tuning.
Interestingly, for $m_{DM} \gsim 200$ GeV, a large portion of the
parameter space (blue points) with $\vert \bar{\lambda}\vert
\gtrsim 0.1$ evade all current experimental constraints as well as the
projected Xenon1T sensitivity, in contrast with the non-composite
Higgs portal scenarios. 

Let us emphasize though that the allowed values of $a_{d1}$ are
already highly constrained by LUX and have good prospects to be
  tested by Xenon1T.  Since in a specific model (where $\mG/\mH$ is
specified), $a_{d1}$ is no longer a free parameter, the allowed region
of parameter space will be strongly constrained by direct DM
searches. For example, in the case of $\mG/\mH = SO(6)\times SO(5)$
discussed below, $a_{d1} = 1/2$ and a large portion of the
  parameter space is already excluded for low values of $F$.  For $F
= 2.5$ TeV, if no DM signal is observed in Xenon1T, there will be most
probably only a small range of DM masses allowed for a given $a_{d1}$
value.

\subsubsection{Specific Case:  $\mG/\mH =SO(6)/SO(5)$}
\label{sec:SO6SO5}

Although we have taken $a_{d1}$ as a free parameter, in a specific
composite model, $a_{d1}$ is completely fixed by the choice of the
coset $\mG/\mH$.  Therefore it is instructive to discuss how the
general results presented above apply to a specific model. The minimal
coset consistent with $\mH \subset SO(4)$ is $\mG/\mH = SO(6)/SO(5)$,
where the 5 pNGBs correspond to the Higgs bi-doublet and the DM
singlet. This case has been discussed in detail in
refs.~\cite{Frigerio:2012uc,UrbanoCDM:14}, so we limit our discussion
to re-interpreting the results obtained previously according to the
general parametrization introduced in eq.~(\ref{singLeff}).  

The $\mG/\mH = SO(6)/SO(5)$ case corresponds to the particular choice
of values:
\begin{equation}
a_{2H} = a_{d1} = 1/2\, .
\end{equation}
Notice that we keep $c_4$, $d_4$, $\lambda_{H6}$, $\lambda_3$
and $\lambda_3'$ as free parameters in our analysis.  Once
$a_{d1}$ and $a_{2H}$ are fixed, we can derive from eq.~(\ref{sDDapp})
a simple expression for the enhancement (or suppression) of
$\sigma_{DMp}$ with respect to its non-composite value:
\begin{equation}
\sigma_{DMp}/\sigma_{DMp}^{NC} = \left(1 \pm r \right)^2
\mbox{, where } r = \frac{2}{\lambda_{\xi h}^{NC}} \frac{m_{DM}^2}{F^2}\, ,
\label{rDef}
\end{equation}
where once again $\lambda_{\xi h}^{NC}$ is the value of the Higgs-DM
coupling in the non-composite case required to produce the correct
relic abundance.  Unlike the general case discussed above, now the
enhancement or suppression factor $r$ is fixed (for a given value of
$m_{DM}$). Hence, instead of spanning a wide range of values,
$\sigma_{DMp}$ now lies in two possible branches, one for the case $1
+ r$ (enhancement) and one for $1 - r$ (suppression).

\begin{figure}[h!]
  \begin{center}
     \includegraphics[width=7.5cm]{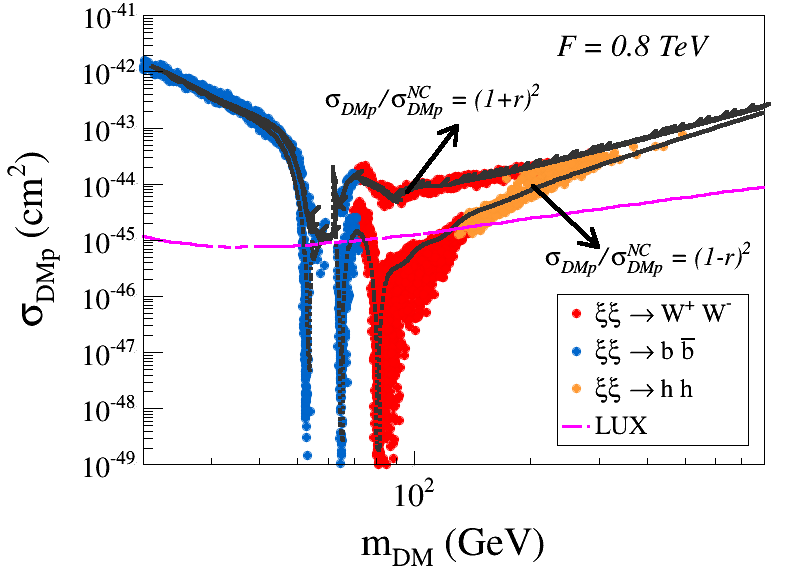}
     \includegraphics[width=7.5cm]{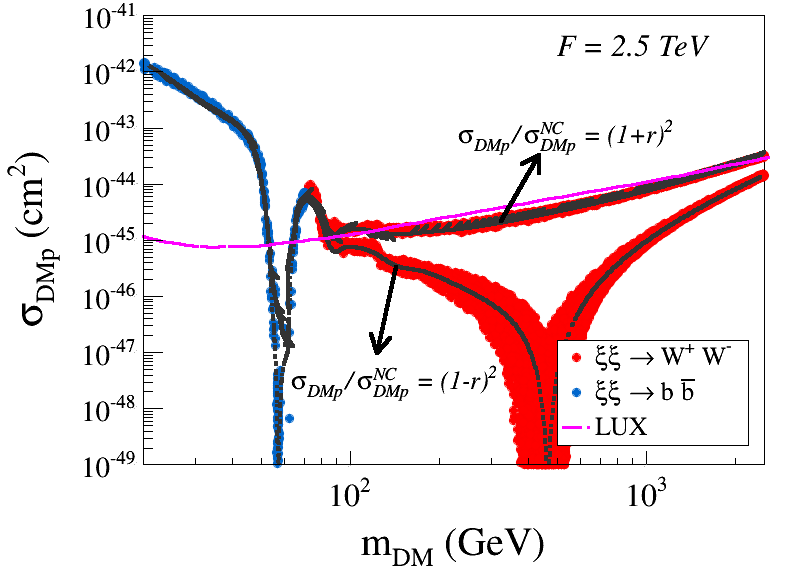}
  \end{center}
  \caption{{\it Singlet DM with $\mG/\mH = SO(6)/SO(5)$:} Values
    of the DM-nucleon scattering cross-section as a function of
    $m_{DM}$. All points satisfy $0.0941 < \Omega_{DM}
    h^2 < 0.127$ while the black continuous lines correspond to
      the approximation of eq.~(\ref{rDef}). We also show with
    different colors the channels contributing with the largest
    branching ratio to the effective annihilation cross-section
    relevant at the time of freeze-out.}
  \label{fig:O6}
\end{figure}

In figure~\ref{fig:O6}, we plot $\sigma_{DMp}$ as
a function of $m_{DM}$, but now for the particular case of $\mG/\mH =
SO(6)/SO(5)$.  We also show with black lines the approximation of
eq.~(\ref{rDef}) for the enhanced ($\sigma_{DMp} = \sigma_{DMp}^{NC}
\left(1 + r \right)^2$) and suppressed ($\sigma_{DMp} =
\sigma_{DMp}^{NC} \left(1 - r \right)^2$) values of the DM-nucleus
cross-section.  As we can see, the solutions obtained through the
  random scan follow closely the approximation of
eq.~(\ref{rDef}).  It is interesting to notice that for $F = 0.8$ TeV all
of the enhanced solutions have already been excluded by LUX, while
only the low mass region ($m_{DM} \lesssim 120$ GeV) of the suppressed
branch is allowed. The situation is however quite different for higher
values of $F$, where we see that most of the parameter space for both
branches has not yet been probed.

Finally we point out that, for $F = 0.8$ TeV, we find no viable
solutions for $m_{DM} > 500$ GeV, even before applying the direct,
indirect and Higgs experimental constraints. Once $m_{DM} > m_h$, the
annihilation into two Higgses can contribute significantly to $\sigv$,
which can be large even if $\lambda_{\xi h}^{DD}\simeq 0$ due to
  derivative coupling between DM and Higgs~\cite{Frigerio:2012uc}. As
a result, the annihilation cross-section becomes too large,
suppressing the relic abundance below the WMAP value.  However this
only happens for small enough values of $F$, as seen in
figure~\ref{fig:O6}.

\subsection{Doublet DM} \label{sec:doubletRes}

If we consider non-minimal cosets, such as e.g. $\mG/\mH =
SO(6)/SO(4)\times SO(2)$ (see section~\ref{sec:SO6SO4SO2} and
e.g. ref.~\cite{Mrazek:2011iu}), it is possible to have non-singlet
representations for the DM multiplet.  Models with multiplet DM differ
greatly from the singlet case in two aspects:
\begin{itemize}
\item they include tree level (unsuppressed) DM-gauge boson   couplings;
\item they allow for co-annihilations between the DM multiplet   components.
\end{itemize} 
Since we require invariance under $SO(4)$, the simplest DM multiplet
representation corresponds to a complex doublet.
In this case, the non-composite limit ($F \to \infty$) corresponds to the
Inert Two Higgs Doublet Model or IDM~\cite{Barbieri:2006dq}.

\subsubsection{Generic composite doublet DM}
\label{sec:gener-comp-doubl}

The generic doublet case Lagrangian corresponds to eq. (\ref{eq:l0}) with the
addition of
\begin{align}
\mathcal{L}^{(2)} & =
 \left(\Dmu \xi\right)^\dagger \DMu \xi 
-  \mu_{\xi}^2 |\xi|^2 - \lambda_3 \left(1 + \frac{\lambda_3'}{F^2}
|H|^2\right) |\xi|^2 |H|^2  - \lambda_4 \left(1 +
\frac{\lambda_4'}{F^2}|H|^2\right) |\xi^\dagger H|^2 \nonumber \\
& - \frac{\lambda_5}{2} \left(1 + \frac{\lambda_5'}{F^2}|H|^2\right)
\left[\left(\xi^\dagger H\right)^2 + \text{h.c.} \right] 
+\frac{a_{d1}}{2F^2}\dmu|H|^2\dMu|\xi|^2 \label{doubLeff0} 
\\
& +\frac{a_{d2}}{F^2}\left(H^\dagger \Dmu \xi + \text{h.c.} \right)\left(\xi^\dagger
\DMu H + \text{h.c.} \right) + \frac{a_{d3}}{F^2} \left[ \dmu \left(\xi^\dagger H +
\text{h.c.}\right) \right]^2  \nonumber \\
& +\frac{a_{d4}}{F^2} \left[ \xi^\dagger \overleftrightarrow{D}_{\mu} \xi
H^\dagger \overleftrightarrow{D}^{\mu} H +  \xi^\dagger
\overleftrightarrow{D}_{\mu} \xi^C H^{C\dagger} \overleftrightarrow{D}^{\mu} H 
- \xi^\dagger \vec{\sigma}
 \overleftrightarrow{D}_{\mu} \xi H^\dagger \vec{\sigma} \overleftrightarrow{D}^{\mu} H
 + \text{h.c.} \right] \nonumber \\
 & - \left[\frac{d_4}{F^2}
 |\xi|^2 \left(y_t \bar{Q}_L H^c t_R + y_b \bar{Q}_L
H b_R\right) + \text{h.c.}\right]  \nonumber \\
 & - \frac{d_6}{F^2} 
  \left[\xi^\dagger \vec{\sigma} \xi \left(y_t
 \bar{Q}_L \vec{\sigma} H^c t_R - y_b \bar{Q}_L \vec{\sigma} H b_R \right) 
+  y_b \xi^{c\dagger} \vec{\sigma} \xi \bar{Q}_L \vec{\sigma} H^c b_R
+ y_t \xi^{\dagger} \vec{\sigma} \xi^c  \bar{Q}_L \vec{\sigma} H t_R + \text{h.c.}
\right]\, , \nonumber
\end{align}
where the coefficients $a_{d1,\ldots,d4},d_4, d_6, \lambda_{3,4,5}'$
are taken to be real ${\cal O}$(1) parameters, the couplings
$\lambda_{3,4,5}$ are allowed to vary in the window $[-4\pi, 4\pi]$
and $\mu_{\xi}$ is the DM bare mass. The Pauli matrices are denoted by
$\vec\sigma=\{\sigma_1,\sigma_2, \sigma_3\}$ and $ \phi^\dag
\overleftrightarrow{D}_{\mu}\psi\equiv \phi^\dag D_{\mu}\psi - (
D_{\mu}\psi)^\dag \phi$. We also define
\begin{equation}
\xi = \frac{1}{\sqrt{2}} \left(\begin{array}{c}
-i H^+ \\ H^0 + i A^0
\end{array}\right)\, ,
\end{equation}
see section~\ref{doubletLag} for more details. After EWSB, the DM multiplet
acquires the following masses:
\begin{eqnarray}
  \label{eq:massesld}
  m_{DM}^2 \equiv m_{H^0}^2   &=& \mu_\xi^2 +\frac{v^2}{2}\left(\lambda_3
  +\lambda_4+\lambda_5+\frac{\lambda_3
  \lambda_3'+\lambda_4 \lambda_4'+\lambda_5 \lambda_5'}{2F^2}v^2\right)\, ,
  \label{eq:mh0}\\
  m_{A^0}^2   &=& \mu_\xi^2 + \frac{v^2}{2}\left(\lambda_3
  +\lambda_4-\lambda_5+\frac{\lambda_3
  \lambda_3'+\lambda_4
  \lambda_4'-\lambda_5\lambda_5'}{2F^2}v^2\right),\label{eq:ma0}\\
  m_{H^\pm}^2 &=& \mu_\xi^2 + \frac{v^2}{2}\lambda_3 \left(1
  +\frac{\lambda_3'}{2F^2}v^2\right).\label{eq:mhc}
 \end{eqnarray}
In the results presented below we consider only the cases where
$m_{H^0} < m_{A^0},m_{H^\pm}$, so $H^0$ is always the
DM.\footnote{However we expect all the results derived below to remain
  valid in the case $A^0$ is the DM particle.} From
eqs.~(\ref{eq:mh0})-(\ref{eq:mhc}) we see that the three masses are in
principle independent. However, large mass splittings are  limited by
the perturbativity requirements on the $\lambda_i,\lambda_i'$
couplings.

Although the inclusion of the dimension-6 operators induced by the
strong sector introduces 9 new coefficients, most of the DM properties
depend on a subset of these (or appropriate linear combinations). For
instance, the Higgs-DM coupling is given by:
\begin{equation}
  \lambda_{\xi h} = \frac{\bar{\lambda}}{2} -\left(a_{d1}+2 a_{d2}+ 4
  a_{d3}\right)\frac{p_h^2}{4 F^2}+ a_{d3}\frac{p_h^2 - 2 m_{DM}^2}{F^2}\, ,
\end{equation}
where $p_h$ is again the Higgs momentum and
\begin{equation}
\bar{\lambda} \equiv \sum_{i=3,5} \lambda_i \left(1 +
  \lambda_i'\frac{v^2}{F^2}\right) \, .
\end{equation}
It is useful to distinguish between the DM-Higgs coupling which enter 
$s$-channel annihilations ($p_h^2 \simeq 4 m_{DM}^2$) and $t$-channel scattering
($p_h^2 \simeq 0$). The first case is relevant to $\sigv$, while the latter
enters the calculation of $\sigma_{DMp}$, hence we define:
\begin{eqnarray}
  \lambda_{\xi h}|_{p_h^2 = 4 m_{DM}^2} & \equiv& \lambda_{\xi h}^{RD} = 
 \frac{\bar{\lambda}}{2} - \left(a_{d1}+2 a_{d2}+ 2
 a_{d3}\right)\frac{m_{DM}^2}{F^2} \, , \nonumber\\
 \lambda_{\xi h}|_{p_h^2 = 0} & \equiv& \lambda_{\xi h}^{DD} = 
 \frac{\bar{\lambda}}{2} - 2 a_{d3}\frac{m_{DM}^2}{F^2} \, .
 \label{lxiDoublet}
\end{eqnarray}
While $\lambda_{\xi h}^{RD}$ is very similar to eq.~(\ref{l3effs}) for
the singlet case, the coupling relevant for direct detection now
contains a contribution from $a_{d3}$. The latter, however, is only
relevant for large DM masses.  Also notice  that the $H^0 H^0 V_\mu V^\mu$
vertex is not modified by the dimension-6 operators
driven by the $a_{d4}$ or $a_{d2}$ coefficients, since the corresponding
operators vanish for $\xi \to (0,H^0)/\sqrt{2}$ and $H \to (0,h +
v)/\sqrt{2}$.

In the large mass region ($m_{DM} \gg m_W$) the annihilation to gauge
bosons is open and become the dominant annihilation channel.  Even in
the pure gauge limit, i.e. when all quartic couplings are set to zero,
$\sigv$ always gets a non-zero contribution from $H^0 H^0 \rightarrow
WW,ZZ$. Let us focus on the analysis of the latter channel, neglecting
the co-annihilations, and the $H^0 H^0 \rightarrow hh, tt$
channels. The annihilation cross-section into gauge bosons in the high
mass region can be approximated by:
\begin{eqnarray}
\sigv|_{H_0H_0\rightarrow VV} & \simeq& \frac{1}{32 \pi m_{DM}^2} \left(2
|\mathcal{M}_T|^2 + |\mathcal{M}_L|^2 \right) \mbox{, where}\nonumber\\
\mathcal{M}_{T}\left(H^0 H^0 \to V_T V_T \right) & \simeq& \frac{g_V^2}{2}\left(1
+
\frac{v^2}{2 m_{DM}^2} \lambda_{\xi h}^{RD}\right) \simeq \frac{g_V^2}{2},
\nonumber
\\
\mathcal{M}_{L}\left(H^0 H^0 \to V_L V_L \right) & \simeq&
\frac{g_V^2}{2} \frac{m_{DM}^2}{m_V^2}\left[\frac{v^2}{m_{DM}^2}
\lambda_{\xi h}^{RD} + \frac{m_{DM}^2}{\bar{m}_{\xi}^2}\left(\frac{\Delta
m_{\xi}^2}{m_{DM}^2} + 2 a_{d2} \frac{v^2}{F^2} \right) \right],
\label{ampL}  
\end{eqnarray}
where $\mathcal{M}_T$ and $\mathcal{M}_L$ are the contributions to the
amplitude from transverse and longitudinal modes respectively and $g_V
= g(g/c_w)$, $\bar{m}_{\xi}^2 \equiv \left(m_{DM}^2 +
m_{H^+(A^0)}^2\right)/2$, $\Delta m_{\xi}^2 \equiv m_{H^+(A^0)}^2 -
m_{DM}^2$, for $V=W(Z)$. 
From the
above expressions we have:
\begin{equation}
\sigv > \frac{1}{16 \pi m_{DM}^2} |\mathcal{M}_T|^2 \simeq
\frac{g_V^2}{32 \pi m_{DM}^2}\, ,
\end{equation}
resulting in too small relic abundances, unless $m_{DM} \gtrsim 500$
GeV.  This is a well known result from the
IDM~\cite{LopezHonorez:2006gr,Hambye:2009pw} and it still holds in the
composite case, since the contributions from the dimension-6 operators
to $\mathcal{M}_T$ are negligible.  Notice that, in contrast with the
non-composite case~\cite{Hambye:2009pw}, for sufficiently large
$m_{DM}$ the annihilation cross-section is driven by the term
proportional to $a_{d2}$.  Let us stress though that, as
  mentioned in section~\ref{compDM}, our effective Lagrangian approach
  is only valid for $m_{DM} \lesssim F$.  

 From eq.~(\ref{ampL}) we see that if
  the $\mathcal{M}_L$ contribution had to be suppressed one would need small
  mass splittings among the components of the DM doublet ($\Delta
  m_{\xi}^2/\bar{m}_{DM}^2 \ll 1$) and the combination
\begin{equation}
\lambda_{\xi h}^{RD} \frac{v^2}{m_{DM}^2} +
2 a_{d2} \frac{v^2}{F^2} 
\label{condLeff}
\end{equation}
should be small. Since $|\lambda_{\xi h}^{RD}|$ can not be made
arbitrarily large (otherwise $H^0 H^0 \to h h$ enhances $\sigv$),
we will see below that the above condition limits the possible
values of $a_{d2}$.

\subsubsection{Scan results and experimental constraints }
\label{sec:scan-results-exper}

For the numerical results presented below we scan over all the parameters in
eq.~(\ref{doubLeff0}) plus the ones in the Higgs effective Lagrangian
(eq.~(\ref{eq:l0})) within the following ranges:
\begin{eqnarray}
  &2 \mbox{ GeV} < m_{DM}< F \, , \nonumber\\
  &m_{DM} < m_{A^0}<F \, , \nonumber\\
  &{\rm max}(70 \mbox{ GeV},m_{DM}) < m_{H^\pm}<F \, , \nonumber\\
  &-4 \pi <\bar{\lambda} <4\pi \, , \nonumber\\
  &10^{-6} <\lambda_{H6} < 1 \, , \nonumber\\
  &-1 <\lambda_3',\lambda_4',\lambda_5' < 1 \, , \nonumber\\
  &-1<c_4, a_{2H}, d_4, d_6<1\, ,  \nonumber\\
  &-1<a_{d1}, a_{d2}, a_{d3}, a_{d4}<1 \, . \nonumber
\label{eq:parm-dsm}
\end{eqnarray}
 Again, we scan logarithmically over $\lambda$'s ranges and linearly
 on all the other parameters range.  Notice that we have conveniently
 replaced $\lambda_1$, $\lambda_2$, $\lambda_3$ and $\mu_\xi^2$ by
 $m_{DM}$, $m_{A^0}$, $m_{H^\pm}$ and $\bar{\lambda}$. We have also
 imposed on the scans that viable models have a stable EW vacuum with
 $\langle \xi \rangle = 0$~\cite{Ginzburg:2010wa} and are consistent
 with LEPII constraints on BSM states, following
 ref.~\cite{Lundstrom:2008ai}.  We require that new charged scalar
 particles $(H^\pm)$ have masses larger than 70
 GeV, see~\cite{Cao:2007rm,Pierce:2007ut,Gustafsson:2010zz}.  Furthermore,
 as it is well known from IDM results~\cite{LopezHonorez:2006gr}, for
 $m_{DM} \gsim 500$ GeV the correct relic abundance can only be
 obtained for small mass splittings within the DM multiplet. In this
 mass region we have imposed that $m_{H^\pm}-m_{DM} < 30$ GeV and
 $m_{A^0}-m_{DM} < 30$ GeV.\footnote{ 
We have checked that the mass splittings obtained in the scans are always below 35 GeV for all $F$ scales assumed here. For the purpose of this work, the range used in the scans ($< 30$ GeV) is sufficient to discuss the composite DM phenomenology.} We can divide the discussion into two mass regions: a low
 mass region for $m_{DM}< m_h$ and a high mass region for
 $m_{DM}>m_h$.

\begin{figure}[h!]
  \begin{center}
     \includegraphics[width=7.5cm]{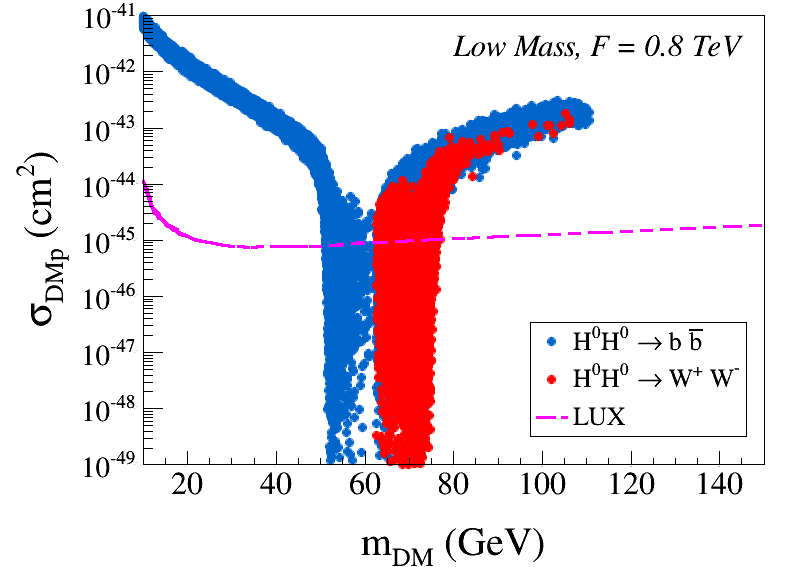}
     \includegraphics[width=7.5cm]{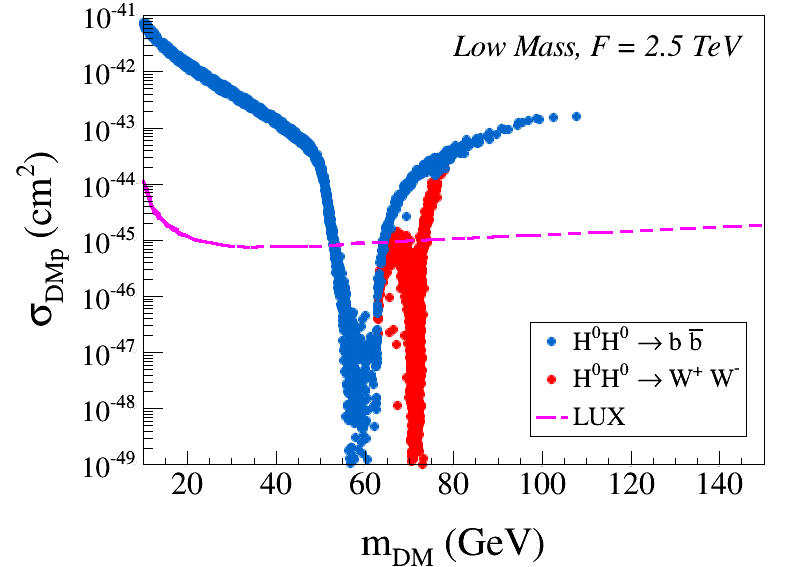}\\
     \includegraphics[width=7.5cm]{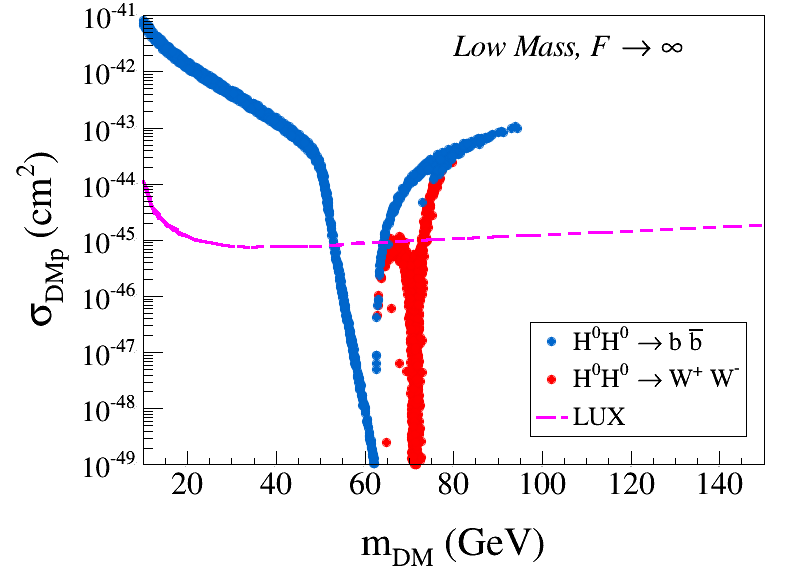}
     \includegraphics[width=7.5cm]{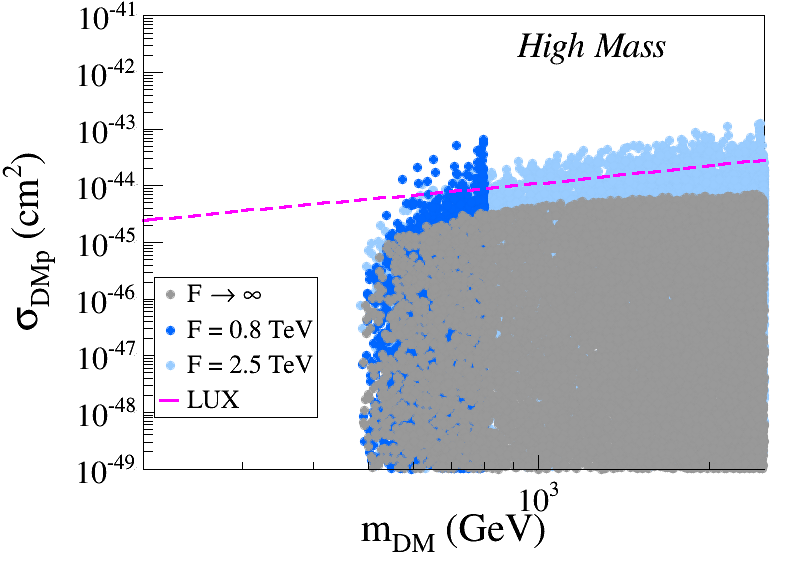}
  \end{center}
  \caption{{\it Generic composite doublet DM:} Values of the
    DM-nucleon scattering cross-section ($\sigma_{DMp}$) as a function
    of $m_{DM}$ satisfying $0.0941 < \Omega_{DM} h^2 < 0.127$. In the top
      plots ($F=0.8$ and 2.5 TeV) and the bottom LH plot
      ($F\rightarrow\infty$), we show the low mass region. The
      channels contributing to the largest branching ratios to the
      effective annihilation cross-section at freeze out are depicted
      with different colors. In
      the bottom RH side plot, we show $\sigma_{DMp}$ in the high mass region for
    $F = 0.8$ TeV (dark blue), $2.5$ TeV (light blue) and the non-composite case ($F \to
    \infty$ in gray) superimposed.}
  \label{fig:doubletSDD}
\end{figure}

In figure~\ref{fig:doubletSDD}, we show the DM-nucleus scattering
cross-section for $F = 0.8$ TeV and 2.5 TeV for $m_{DM}< m_h$ in the
top plots. For reference, we also show the corresponding non-composite
limit, $F\rightarrow \infty$, in the bottom LH plot. In this region of
the parameter space, LEPII~\cite{Cao:2007rm,Lundstrom:2008ai}
constraints require a large mass splitting between $H^0$ and $A^0$ or
$H^\pm$. In particular, all the region with $m_{DM} + m_{A^0} < m_Z$
is excluded due to the constraints on the $Z$-invisible width. This
combined with the lower bound on $m_{H^\pm}$ makes co-annihilations
negligible for $m_{DM} \lesssim 60$ GeV. Since annihilations to gauge
bosons are kinematically forbidden in this region, the main features
distinguishing the doublet and singlet cases -- co-annihilations and
DM gauge couplings -- are negligible for small DM masses and the
doublet DM behaves effectively as singlet DM.

Once $m_{DM} \gtrsim m_W$, the $H^0 H^0 \to V V$ process rapidly
enhances the DM annihilation cross-section.\footnote{Since we include
  the effects of annihilation to off-shell $W$'s and $Z$'s ($H^0 H^0
  \to V^* V$), this channel becomes relevant for slightly smaller
  values of the DM mass ($m_{DM} \gtrsim 60$ GeV), as seen in
  figure~\ref{fig:doubletSDD}.}  For $m_{DM} \lesssim 110$ GeV this
enhancement is still partially suppressed by kinematics and can
result in the correct relic abundance if there is destructive
interference between the $H^0 H^0 \to h^* \to V V$ and $H^0 H^0 \to V
V$ amplitudes~\cite{LopezHonorez:2010tb}, which happens for
$\lambda_{\xi h}^{RD} < 0$.  For larger values of the DM mass, the
contribution from annihilation into gauge bosons is no longer
suppressed and generates too small relic abundances, independently of
the choice of parameters.  This is a well known feature of the IDM and
remains valid in the composite scenario. Nonetheless, in the region
where $H^0 H^0 \to V V$ is (partially) suppressed, the contributions
from the dimension-6 operators to $\lambda_{\xi h}^{RD}$ allow for a
wider range of $\bar{\lambda}$ values consistent with $\Omega_{DM} h^2
\simeq 0.12$.  As a consequence we obtain larger viable
  parameter space in figure~\ref{fig:doubletSDD} for $F=0.8$ TeV and
$F = 2.5$ TeV (top plots), when compared to the non-composite case
(bottom LH plot). Also, the allowed low mass region extends to
slightly higher values of $m_{DM}$ due to the presence of the
dimension-6 operators.

\begin{figure}[h!]
  \begin{center}
     \includegraphics[width=7.5cm]{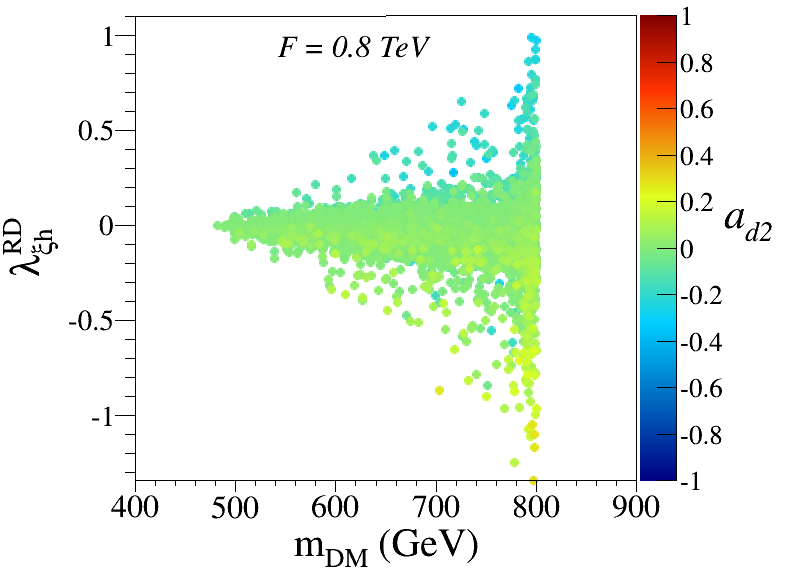}
     \includegraphics[width=7.5cm]{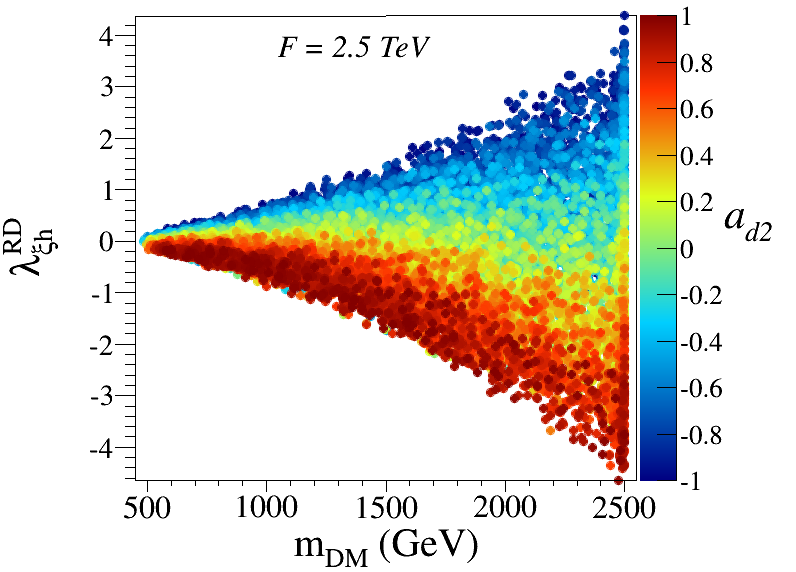}
  \end{center}
  \caption{{\it Generic composite doublet DM in the high mass
      range:} Values for the dimension-6 coefficient $a_{d2}$ in the
    $\lambda_{\xi h}^{RD}$-$m_{DM}$ plane for $F = 0.8$ TeV (left) and
    2.5 TeV (right).}\label{fig:doubletad2}
\end{figure}

In the bottom RH side plot of figure~\ref{fig:doubletSDD} we show
values of $\sigma_{DMp}$ in the high mass region. We see that viable
solutions are only obtained for $m_{DM} \gsim 500$ GeV. The relative
mass splittings between the dark components has to be small.
Typically we obtain $\Delta m_\xi<$ 20 GeV $\ll m_{DM}$. Since
$a_{d2}$ can take negative values, the allowed mass splittings in the
composite case are slightly larger than in the IDM, due to the
cancellation between the $a_{d2}$ and $\Delta m_{\xi}^2$ terms in
eq.~(\ref{ampL}). In addition, in order to properly quantify the
correlations between $\lambda_{\xi h}^{RD}$ and $a_{d2}$, we show in
figure~\ref{fig:doubletad2} the values of $a_{d2}$ with gradient
colors in the $\lambda_{\xi h}^{RD}-m_{DM}$ plane for $F = 0.8$ and
2.5 TeV. We see that $a_{d2}$ and $\lambda_{\xi h}^{RD}$ typically
have opposite signs guaranteeing a suppression of the combination
(\ref{condLeff}). We thus see that all the contributions to ${\cal
  M}_L$ need to be suppressed in order to account for the right relic
abundance.  For $F = 0.8$ TeV, all points have $|a_{d2}|\lesssim 0.3$,
while for $F = 2.5$ TeV $|a_{d2}|$ can be as large as 1, due to the suppression
factor $v^2/F^2 \simeq 0.01$.

From figure \ref{fig:doubletSDD} we see that $\sigma_{DMp}$ can take
larger values in composite scenarios ($F=0.8$ or 2.5 TeV) than in the
non-composite case ($F\rightarrow \infty$). This is particularly
visible in the large mass range in the bottom RH side plot where we
have $\sigma_{DMp} \lsim 10^{-44}\; (10^{-43})$ cm$^2$ for $F \to
\infty$ ($F = 0.8$ TeV). This is easily understood looking at
eq.~(\ref{lxiDoublet}). In the non-composite case $\lambda_{\xi
  h}^{RD} = \lambda_{\xi h}^{DD} = \bar{\lambda}/2$ and $\sigma_{DMp}
\propto \bar{\lambda}^2$, while in composite scenarios we have
\begin{equation}
 \lambda_{\xi h}^{RD} - \lambda_{\xi h}^{DD}  = - \left(a_{d1}+2 a_{d2}\right)\frac{m_{DM}^2}{F^2}\,.
\end{equation}
As a result, $|\lambda_{\xi h}^{DD}|$, the quartic coupling driving
$\sigma_{DMp}$, can get extra contributions due to the $a_{di}$ terms
in composite models.

\begin{figure}[h!]
  \begin{center}
     \includegraphics[width=7.5cm]{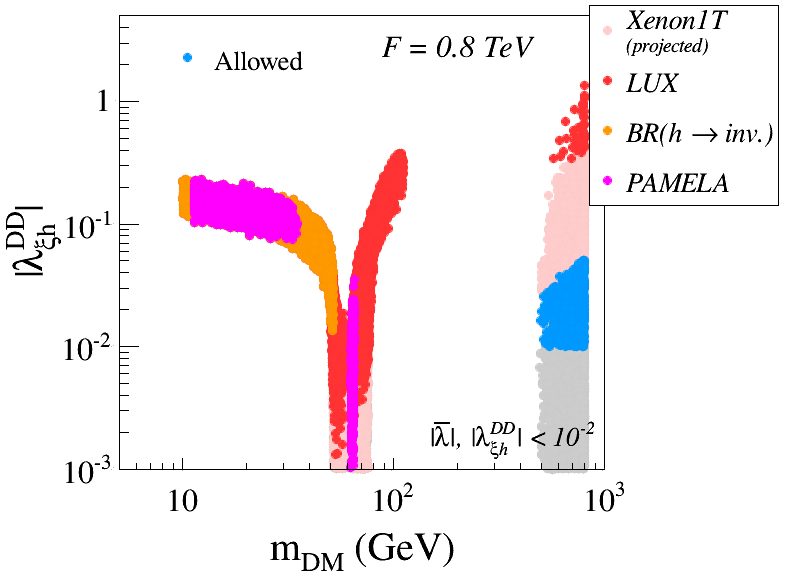}
     \includegraphics[width=7.5cm]{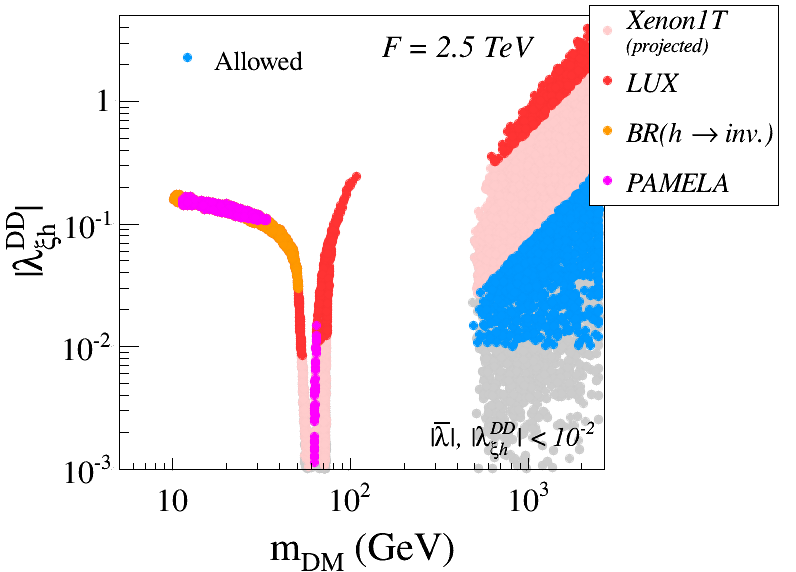}
 \end{center}
  \caption{{\it Generic composite doublet
    DM:} Experimental constraints in the $|\lambda_{\xi h}^{DD}|-m_{DM}$ plane for
    $F =0.8$ TeV (left) and 2.5 TeV (right). See text for details.} \label{fig:constDoublet}
\end{figure}

Figure~\ref{fig:constDoublet} illustrates the experimental constraints
on the generic composite doublet case, the color code is the same as
the one used in section~\ref{sec:exper-constr-s}.  Once again, points
with $|\bar{\lambda}| < 0.01$ are theoretically disfavored in
realistic composite models, where we expect
$\mathcal{O}\left(\lambda_1\right) \simeq
\mathcal{O}\left(\lambda_i\right) \sim 0.1$. Similarly, unless the
$a_{d3}$ and $\bar{\lambda}$ contributions cancel each other in
eq.~(\ref{lxiDoublet}), at the price of some fine tuning, we will
typically have $\mathcal{O}\left(\bar{\lambda}\right) \sim
\mathcal{O}\left(\lambda_{\xi h}^{DD} \right)$ and too low values of
$\lambda_{\xi h}^{DD}$ will also be disfavored. Hence, we consider all
points with $\rm min(|\lambda_{\xi h}^{DD}|,|\bar{\lambda}|) <
10^{-2}$ as theoretically disfavored and they are shown in gray.  As
we can see, in the low mass region all points are either excluded by
current experimental constraints or involve too small values of $\vert
\lambda_{\xi h}^{DD}\vert, \vert\bar{\lambda}\vert$ to be
theoretically acceptable, except for a small number of points
  (with $\rm min(|\lambda_{\xi h}^{DD}|,|\bar{\lambda}|) > 10^{-2}$)
  lying between the resonance region and the $W$ threshold for large
  values of $F$ . Therefore, most of the low mass region of generic
  composite doublet DM models is already experimentally excluded or
  theoretically disfavored. In contrast, the high mass region has
just been marginally probed by LUX and it even Xenon1T will not be
able to test it in its entirety.
\subsubsection{Specific Case: $\mG/\mH =SO(6)/SO(4)\times SO(2)$}
\label{sec:SO6SO4SO2}

Once again it is interesting to apply the general results presented above
to a specific model. Here we take $\mG/\mH = SO(6)/SO(4)\times
SO(2)$, where the only pNGBs are the Higgs bi-doublet and the complex
doublet DM.
This case corresponds to the effective Lagrangians in eqs.~(\ref{eq:l0}) and
(\ref{doubLeff0})  with:
\begin{align}
a_{2H} & = \frac{1}{2} \mbox{, } a_{d2} = 1 \mbox{ and } a_{d1} = a_{d3} = a_{d4} =
0\, .
\end{align}
Notice that, according to the results of
figure~\ref{fig:constDoublet}, we do not expect many  allowed
solutions in the low mass range after the LUX constraints have been
imposed for $|\bar{\lambda}| > 10^{-2}$. Therefore here we only
discuss the high mass ($m_{DM} > 500$ GeV) solutions.

\begin{figure}[h!t]
  \begin{center}
  \includegraphics[width=7.5cm]{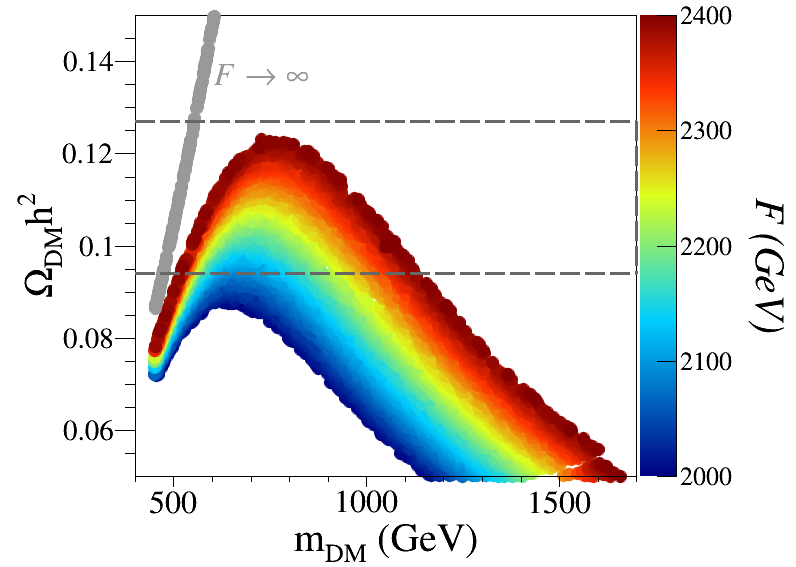}
  \includegraphics[width=7.5cm]{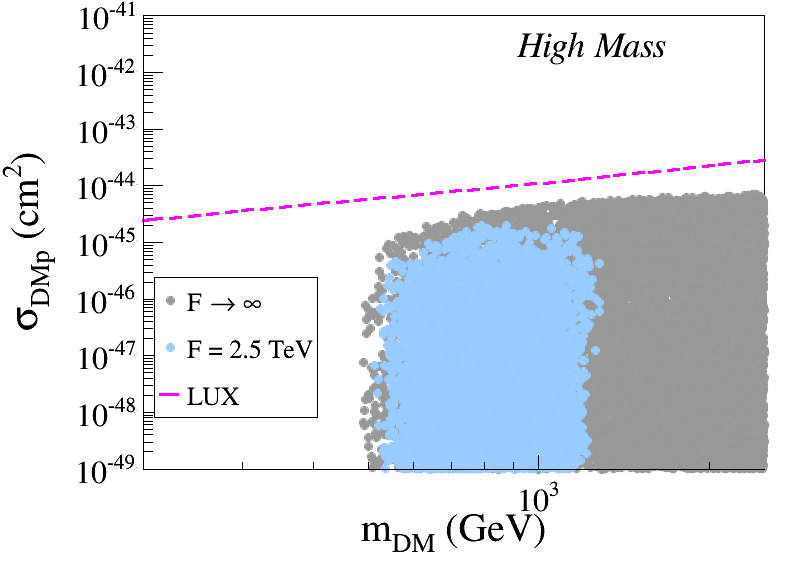}
  \end{center}
  \caption{{\it Doublet DM with $\mG/\mH = SO(6)/SO(4)\times
        SO(2).$} {\it Left}: Values of the DM relic density as a
    function of $m_{DM}$ for 2.0 TeV $< F < 2.4$ TeV fixing
    $\bar{\lambda} = d_4 = d_6 = \lambda_i' = 0$ and $m_{A^0} =
    m_{H^\pm} = m_{DM}$. The gray points correspond to the
    non-composite case with $\bar{\lambda} = 0$. We also show
    by dashed lines the $2\sigma$ CMB bounds on
    $\Omega_{DM} h ^2$. {\it Right}: $\sigma_{DMp}$ as a function of
    $m_{DM}$ for points in agreement with the CMB bounds for
      $F=2.5$ TeV and $F\rightarrow \infty$ leaving all quartic and
      derivative couplings as free parameters.}
  \label{fig:fscan}  
\end{figure}

One important consequence from specifying the $\mG/\mH$ coset is that
it fixes the value of $a_{d2}$. As we have seen above, the high mass
region requires either large values of $F$ or small values of $a_{d2}$
(see figure \ref{fig:doubletad2}).  Hence, for $\mG/\mH =
SO(6)/SO(4)\times SO(2)$, where $a_{d2} = 1$, we do not expect viable
solutions for large DM masses and small values of $F$.  This is
highlighted in the LH of figure~\ref{fig:fscan}, where we allow
$\Omega_{DM} h^2$ to vary outside the $0.09 \lesssim \Omega_{DM} h ^2
\lesssim 0.12$ range and show the relic density as a function of
$m_{DM}$ for different $F$ values, shown by the color gradient.  For
this plot, we take $m_{A^0} = m_{H^\pm} = m_{DM}$ and $\bar{\lambda} =
\lambda_i' = 0$ since this choice approximately maximizes (minimize)
the relic abundance ($\sigv$) and we take $d_4=d_6=0$, since these
coefficients play no major role in the high mass region.

We can understand the behavior of $\Omega_{DM} h ^2$ in the LH of
figure~\ref{fig:fscan} by analyzing the $m_{DM}$ and $F$ dependence of
$\sigv(\xi\xi\rightarrow VV)$ derived in eq.~(\ref{ampL}), but in the
``pure gauge limit'' (all quartic couplings set to zero).  The relic
abundance first grows with $m_{DM}$ due to the usual dependence
$\Omega_{DM} h ^2\propto 1/\sigv\propto m_{DM}^2$.  This is indeed
what we observe for the lowest range of masses in
figure~\ref{fig:fscan} (LH side).  The non-composite case in the same
limit is plotted for reference with gray points and also show this
growing dependence in $m_{DM}$.  When $m_{DM}$ becomes comparable to
$F$, the dimension-6 derivative operator starts to dominate $\sigv$
and the contribution from the $a_{d2}$ term results in $\Omega_{DM} h
^2\propto 1/\sigv\propto 1/m_{DM}^2$. For a given $F$ scale there is
thus a fixed mass range, bounded from below and above, of viable DM
candidates, i.e. in agreement with the CMB bound depicted with
vertical dashed gray lines in figure~\ref{fig:fscan} (LH side).  This
viable mass range widen with increasing $F$. Indeed, we see that for
e.g. $F = 2.4$ TeV we have 550 GeV $\lesssim m_{DM} \lesssim 1.2$ TeV,
while for e.g.  $F= 2.1$ TeV we have 600 GeV $\lesssim m_{DM} \lesssim
800$ GeV. In addition, once $F < 2$ TeV, $\Omega_{DM} h^2 < 0.09$ for
any value of $m_{DM}$ in the large mass range. Therefore, for viable
composite scenarios with $\mG/\mH = SO(6)/SO(4)\times SO(2)$ one needs
to consider $F > 2$ TeV.

 In figure~\ref{fig:fscan} (RH side) we show results for a scan
 on the parameter space of  $\mG/\mH = SO(6)/SO(4)\times SO(2)$, where the
 non-derivative coefficients are still free parameters.
 We have explicitly verified that even when all non-derivative parameters
 are allowed to vary there are no viable solutions for $F = 0.8$ TeV.
 We thus present points only for $F = 2.5$ TeV and
  $F\rightarrow \infty$. As shown by the results, for $F = 2.5$ TeV, we only
  obtain the correct relic abundance for 500 GeV $\lesssim m_{DM} \lesssim 1.1$
  TeV, as expected from our discussion above.

\subsection{Triplet DM and Higher Representations}\label{sec:tripletRes}

In sections~\ref{sec:singletRes} and \ref{sec:doubletRes} we have
discussed the cases of the singlet and complex doublet DM, which can
be realized for specific choices of the coset $\mG/\mH$, such as the
$SO(6)/SO(5)$ and $SO(6)/SO(4)\times SO(2)$ ones discussed
above. If one wants to consider even larger symmetry groups
  $\mG$, it is possible that in the low energy effective theory one
ends up with a triplet DM or even higher $SU(2)_L$ representations. As
it is well known from non-composite
models~\cite{Cirelli:2005uq,Hambye:2009pw}, such cases are highly
constrained and tend to require multi-TeV DM masses.  In this section
we will discuss how these features are affected by the inclusion of
the dimension-6 operators induced by the strong sector.  As discussed
in section~\ref{sec:results}, we focus here on real representations,
which correspond to $\xi^C = \xi$ and simplify considerably the
analysis. Also, after EWSB, the lightest component of the DM
  candidate from a complex representation usually is a charged field
  and do not provide a viable DM candidate  \cite{Hambye:2009pw}.  

The effective DM Lagrangian for any real
DM multiplet is given by:
\begin{eqnarray}
\mathcal{L}^{(2)} & =&  \left(\Dmu \xi\right)^\dagger \DMu \xi 
-  \mu_{\xi}^2 |\xi|^2 - \lambda_3 \left(1 + \frac{\lambda_3'}{F^2}
|H|^2\right) |\xi|^2 |H|^2 \nonumber \\
& &- \frac{\lambda_4}{F^2} \xi^\dagger \left\{ \Gamma^i,\Gamma^j \right\} \xi
H^\dagger \sigma^i H H^\dagger \sigma^j H
- \frac{\lambda_5}{F^2} \xi^\dagger \left\{ \Gamma^i,\Gamma^j \right\} \xi
H^{c\dagger} \sigma^i H H^\dagger \sigma^j H^c \label{tripLeff0}
 \\
& &+ \frac{a_{d1}}{2F^2}\dmu |\xi|^2 \dMu |H|^2
- \frac{a_{d4}}{F^2} \xi^\dagger \vec{\Gamma}
\overleftrightarrow{D}^\mu \xi H^\dagger \vec{\sigma} \overleftrightarrow{D}_\mu
H
\nonumber \\
 & &- \frac{d_4}{F^2}|\xi|^2 \left(y_t \bar{Q}_L H^c t_R + y_b \bar{Q}_L H
 b_R  + \text{h.c.}\right) \, , \nonumber 
\end{eqnarray}
where the coefficients $a_{d1,d4}$, $d_4$ and $\lambda_{3,4,5}'$ are taken
to be real ${\cal O}$(1) parameters, the couplings $\lambda_i$ with $i=3,4,5$
 are allowed to vary in the window
  $[-4\pi, 4\pi]$ and $\mu_{\xi}$ is the DM bare mass. The Pauli
  matrices are again denoted by $\vec\sigma=\{\sigma_1,\sigma_2,
  \sigma_3\}$ while the $\Gamma$ matrices are the generators of the
  representation $n$ of SU(2)$_L$ satisfying $[\Gamma_i,\Gamma_j]=i
  \epsilon_{ijk}\Gamma_k$ (see appendix~\ref{tripletLag} for more details).

\subsubsection{Generic composite triplet DM}
\label{sec:gener-comp-tripl}

In order to provide some insight on the general behavior of a
composite DM multiplet of dimension $n>2$ of SU(2)$_L$ we first focus on
the case of the triplet. Later we will discuss how these results
generalize to larger representations. In the triplet case, we use the
following generators (in the spherical basis):
\begin{equation}
\label{eq:generatortripletspheric}
\Gamma_1=\frac{1}{\sqrt{2}}\begin{pmatrix} 0 & -1 & 0 \\ -1 & 0 & 1 \\
0 & 1 & 0
\end{pmatrix} \, , \quad
\Gamma_2=\frac{1}{\sqrt{2}}\begin{pmatrix} 0 & i & 0 \\ -i & 0 & -i \\
0 & i & 0  \end{pmatrix} \, , \quad
\Gamma_3=\begin{pmatrix} 1 & 0 & 0 \\ 0 & 0 & 0 \\ 0 & 0 & -1 \end{pmatrix}~.\\  
\end{equation}
and $\xi=(T^+, T^0,T^-)^T$.  Using eq.~(\ref{tripLeff0}) with the
above representation for the $\Gamma$ matrices, we obtain the
following scalar mass spectrum:
 \begin{eqnarray}
    m^2_{DM} \equiv m^2_{T^0}&=&\mu_{\xi}^2 +v^2 \left(\frac12 \lambda_3+\frac14
    \lambda_3 \lambda_3'\frac{v^2}{F^2}+
    \lambda_5\frac{v^2}{F^2}\right) \, , \label{eq:mT0}\\
   m^2_{T^\pm}&=&\mu_{\xi}^2 +v^2 \left(\frac12 \lambda_3+\frac14
   \lambda_3 \lambda_3'\frac{v^2}{F^2}+ \frac12
   (\lambda_4+\lambda_5)\frac{v^2}{F^2}\right) \, .
\label{eq:mTc}
 \end{eqnarray}
Notice that in the
triplet case, once the values of $m_{DM}$, $\lambda_4$ and $\lambda_5$
are given, the value of $m_{T^\pm }$ is fixed.  

In the non-composite case ($F \to \infty$), the triplet is exactly
degenerate (except for small EW loop corrections,
see~\cite{Cirelli:2005uq}). In the latter case, either LEP constraints
excluding the existence of low mass new charged particles or too
strong annihilation into gauge (virtual) bosons\footnote{In the
  triplet case $T^0 T^0 \to V V^{*}$ is already dominant for $m_{DM}
  \gtrsim 50$ GeV.} prevent the existence of viable DM candidate in
the low mass range (i.e. for $m_{DM}<m_h$).  In the composite
scenario, however, this is no longer the case since the dimension-6
operators induce a mass splitting, given by:
\begin{equation}
m_{T^\pm}^2 - m_{DM}^2  = \frac{v^4}{2 F^2}\left(\lambda_4 - \lambda_5 \right) \, .
\end{equation}
Therefore, assuming $|\lambda_{4,5}| \lesssim 1$, we can have $m_{DM}<
60$ GeV imposing $m_{T^\pm} > 70$ GeV, if $F \lesssim 1.5$ TeV. Larger
values of $F$, however, will not generate sufficient mass splitting to
allow for viable solutions in the low mass region. We will analyze
this in more details in section~\ref{sec:scan-results-triplet}.

The effective Higgs-DM coupling in the triplet case is simply:
\begin{eqnarray}
  &\lambda_{\xi h}& =  \frac{\bar{\lambda}}{2} -
  a_{d1}\frac{p_h^2}{4 F^2}\; ,\label{eq:lxihT}\\
&\mbox{where}& \bar{\lambda} = \lambda_3 \left(1 + \lambda_3'\frac{v^2}{F^2}\right) + 4
\lambda_5 \frac{v^2}{F^2}\, .\label{eq:lbarT}
\end{eqnarray}
 Unlike the doublet case, the DM-gauge boson couplings are only
 affected by one single coefficient ($a_{d4}$) in
 eq.~(\ref{tripLeff0}).  While the contributions to the quartic
 vertices ($\xi^2 V_\mu V^\mu$) are usually negligible:
\begin{equation}
\mathcal{L}_{\rm triplet} \supset 2 g^2 \left(1 -
a_{d4} \frac{v^2}{F^2}\right) T^0 T^0 W^+ W^- \, ,
\end{equation}
  as they are always suppressed by $v^2/F^2$, the triple vertex
  couplings ($\xi^2 V_\mu$) can be enhanced at large $m_{DM}$.  In
 addition let us emphasize that, for the triplet and higher
 representations, the coupling to the gauge bosons is always larger
 than in the doublet case because it scales like $(n^2-1)^2$.  In the
 case of the doublet, we saw that once $m_{DM} \simeq m_W$, the
 (co-)annihilation into gauge bosons generate too large $\sigv$,
 resulting in too low relic densities until $m_{DM} \gtrsim 500$ GeV.
 This feature is only enhanced in the triplet case, since its
 couplings to gauge bosons are larger, as it is well known in the
 non-composite case~\cite{Cirelli:2005uq,Hambye:2009pw}.  We also
 point out that, due to $CP$-invariance, $Z$ decays to DM pairs are
 always forbidden, so the $Z$-width constraints do not apply to
 the case of real triplets.

\subsubsection{Scan results and experimental constraints}
\label{sec:scan-results-triplet}

In order to fully cover the allowed parameter space in a model independent
approach, we once again scan over the effective Lagrangian couplings within the
following ranges:
\begin{eqnarray} &20 \mbox{ GeV} < m_{DM}< F \, , \nonumber\\
  &10^{-6} <\lambda_{H6} < 1 \, , \nonumber\\
  &-4\pi <\lambda_3 < 4 \pi \, , \nonumber\\
  &-1 <\lambda_3',\lambda_4,\lambda_5 < 1 \, , \nonumber\\
  &-1<c_4, a_{2H}, d_4<1 \, , \nonumber\\  
  &-1<a_{d1}, a_{d4}<1 \, . \nonumber
\label{eq:parm-tsm}
\end{eqnarray}
Again, we scan logarithmically over $\lambda$'s ranges and linearly
 on all the other parameters.

\begin{figure}[h!]
  \begin{center}    
     \includegraphics[width=7.5cm]{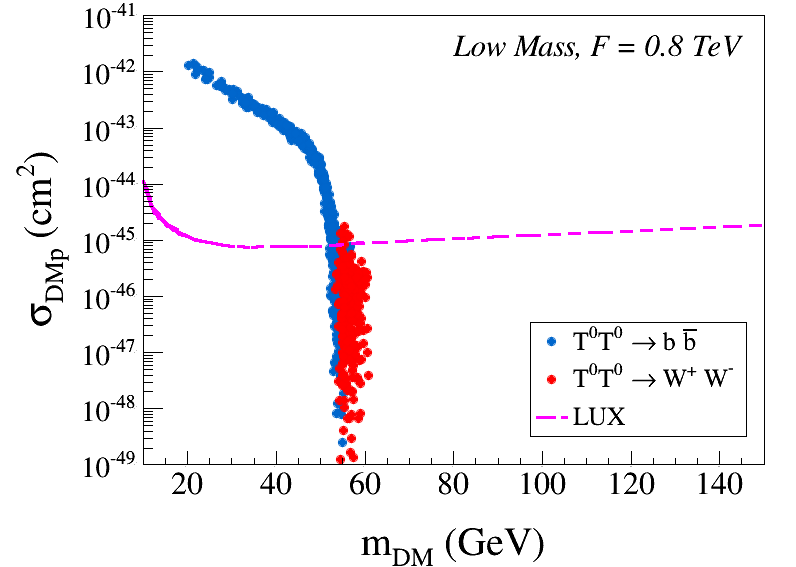}
     \includegraphics[width=7.5cm]{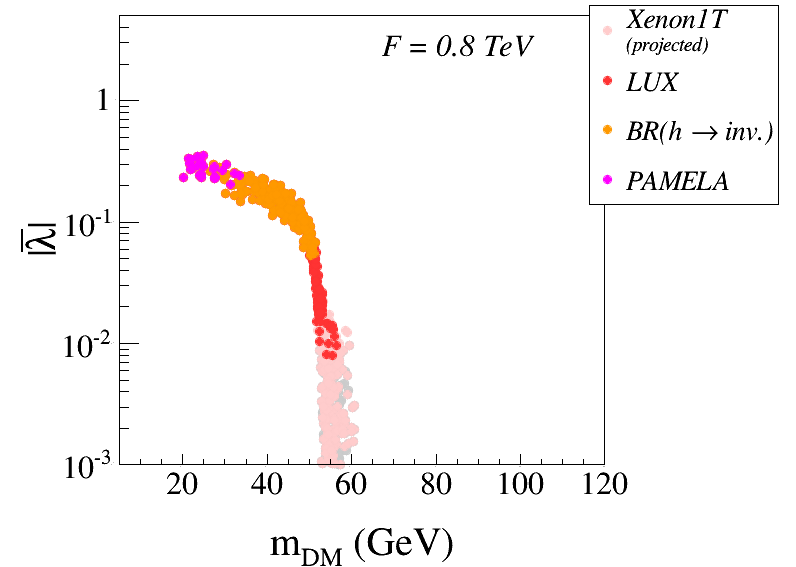}     
  \end{center}
  \caption{{\it Generic composite triplet DM in the low mass
        range.} {\it Left}: Values of the DM-nucleon scattering
    cross-section as a function of $m_{DM}$ satisfying
    to $0.0941 < \Omega_{DM} h^2 < 0.127$. The channels contributing with
      the largest branching ratio to the annihilation cross-section at
      the time of freeze-out are depicted with different colors.
    {\it Right}: Experimental constraints on the low mass region. See
    text for details.}
  \label{fig:tripletLM}
\end{figure}

The results of these scans for the low mass region are shown in the LH
side of figure~\ref{fig:tripletLM}, where we show the mass splitting
and the DM-nucleus scattering cross-section as a function of
$m_{DM}$. We take $F = 0.8$ TeV, since higher values of $F$ will only
reduce the allowed region, due to the suppression of the mass
splitting between the neutral and the charged component. We also show
by blue (red) points the region where $T^0 T^0 \to \bar{b} b$ ($T^0
T^0 \to W^{(*)} W$) dominates the annihilation cross-section in the
early universe.  We see that, for $m_{DM} \simeq 50 $ GeV,
annihilations to (off-shell) gauge bosons become dominant and, for
$m_{DM} \gtrsim 60 $ GeV, $\sigv$ becomes too large, resulting in
relic abundances below the CMB bounds.  In the RH side of
figure~\ref{fig:tripletLM} we show the experimental constraints on the
low mass region.  As we can see, most of the parameter space is
  excluded by experimental constraints, or theoretically disfavored
  for extremely low values of $\bar{\lambda}$ ($<10^{-2}$). Only a
  small number of points within Higgs-resonance and $W$ threshold and
  with $\bar{\lambda}>10^{-2}$ are still viable and will eventually be
  tested by Xenon1T experiment.  We conclude that, despite allowing
for large mass splittings, the composite triplet scenario, alike the
non-composite case, remains mostly excluded in the low mass
region.

In a way analogous to the doublet scenario, for $m_{DM} \gg m_W$, the
annihilation cross-section  into gauge bosons becomes suppressed
  enough to give rise to the right relic abundance. In this high mass
  region, it takes the following form:
\begin{align}
\sigv|_{T_0T_0\rightarrow VV} & \simeq \frac{1}{32 \pi m_{DM}^2}\left(2 |\mathcal{M}_T|^2 +
|\mathcal{M}_L|^2 \right) \mbox{, where}\nonumber\\
\mathcal{M}_{T}\left(T^0 T^0 \to V_T V_T \right) & \simeq
2 g_V^2 \left[ 1 - a_{d4}\frac{v^2}{F^2} +
\frac{\bar{\lambda}}{8} \frac{v^2}{m_{DM}^2} - a_{d1}
\frac{v^2}{2 F^2} m_{DM}^2 \right] \nonumber \\
& \simeq 2 g_V^2 \left( 1 - a_{d4}\frac{v^2}{F^2} + \frac{v^2}{2
m_{DM}^2}\lambda_{\xi h}\right)\; , \nonumber \\
\mathcal{M}_{L}\left(T^0 T^0 \to V_L V_L \right) & \simeq
\frac{g_V^2}{2}\frac{v^2}{m_V^2} \lambda_{\xi h} \simeq
\lambda_{\xi h} \, , 
\label{ampTtrip}
\end{align}
where we have taken $m_{DM} \simeq m_{T^+}$, since the small mass
splitting can be neglected in the high mass region.  Notice that
  in the large mass regime, we typically have $\lambda_{\xi h} v^2/(2
  m_{DM}^2)\ll 1$, and for the $F \gg v$ considered here we also have
  $a_{d4} v^2/F^2 < 1$. These two contributions to ${\cal M}_T$
  will thus typically be small.

\begin{figure}[h!]
  \begin{center}
     \includegraphics[width=7.5cm]{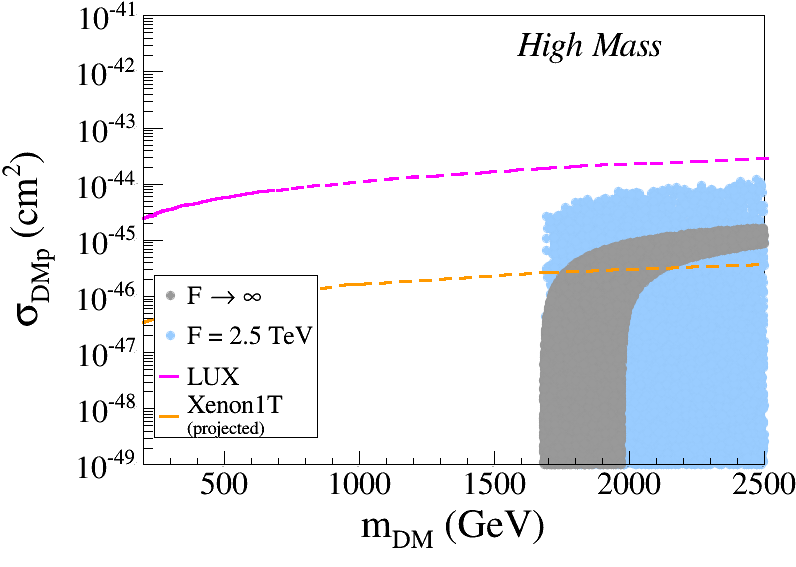}
     \includegraphics[width=7.5cm]{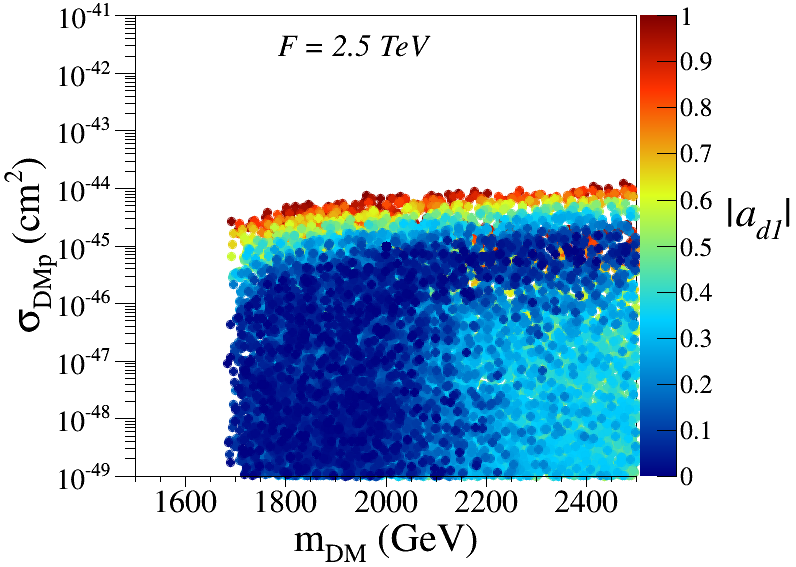}     
  \end{center}
  \caption{ {\it Generic composite triplet DM in the large mass
        range:} Values of the DM-nucleon scattering cross-section as
    a function of $m_{DM}$ satisfying to $0.0941 < \Omega_{DM} h^2 < 0.127$. {\it Right}:
    Values of $|a_{d1}|$ in the $\sigma_{DMp}-m_{DM}$ plane for $F =
    2.5$ TeV.}
  \label{fig:tripHM}
\end{figure}

In the LH side of figure~\ref{fig:tripHM} we show the high mass region
results for $F = 2.5$ TeV and the non-composite case ($F \to
\infty$). As we can see, the lowest $m_{DM}$ values which generate the
correct relic abundance are $m_{DM} \gtrsim 1.7$ TeV. Hence there are
no possible solutions for $F = 0.8$ TeV, since our effective
Lagrangian assumes $m_{DM} < F$. We also show the LUX and projected
Xenon1T constraints. As we can see, the large mass region has not yet
been probed by LUX and will only be partially tested by Xenon1T. From
this plot, we see that while in the non-composite case a
non-observation of signal in Xenon1T would imply $m_{DM} \lesssim 2.1$
TeV, in the composite scenario this upper bound can be easily avoided
due to the presence of the dimension-6 derivative operator.  

Let us emphasize that for such large triplet masses the Sommerfeld
effect can be non-negligible~\cite{Cirelli:2007xd}. A full treatment
of the Sommerfeld enhancement within composite dark matter scenarios
is however beyond the scope of this work. Let us mention though that,
already in the non-composite case, one expects a non-perturbative
enhancement factor of the effective coupling driving the annihilation
cross section which is approximately constant away from the resonances
and is about 1.6 in the triplet
case~\cite{Cirelli:2007xd,Hambye:2009pw}. This  implies an even
stronger lower bounds on $m_{DM}$ in the high mass region. For
instance, if the Sommerfeld enhancement~\cite{Sommerfeld:1931} is
included in the non-composite limit, one would obtain $m_{DM} \gtrsim
2.1$ TeV, instead of $1.7$ TeV.

The effective Lagrangian in the non-composite limit contains only two
parameters ($\lambda_3$ and $m_{DM}$) and, as in the singlet case, the
DM-nucleus cross-section value is fixed (for a given $m_{DM}$) once
the constraint on the relic abundance is imposed. The band seen in the
LH side of figure~\ref{fig:tripHM} for $F \to \infty$ is simply due to
the fact that we allow $\Omega_{DM} h^2$ to vary within the interval
$[0.0941,0.127]$.  For the composite case, however, $\lambda_{\xi h}$
receives contributions from the dimension-6 coefficients $\lambda_3'$
and $a_{d1}$. In this case, we can have larger or smaller values of
$\sigma_{DMp}$ (when compared to the non-composite case), depending on
the values of $a_{d1}$. In the RH side of figure~\ref{fig:tripHM} we
show the values of $|a_{d1}|$ in the $\sigma_{DMp}-m_{DM}$ plane.  We
can clearly see that values of $|a_{d1}| \gtrsim 0.3$ correspond to
the regions where the DM-nucleus cross-section is either suppressed or
enhanced with respect to the non-composite values.  It is also
interesting to notice that, while the low mass region requires $F
\lesssim 1.2$ TeV in order to generate enough mass splitting, the high
mass region requires $F > m_{DM} \gtrsim 1.7$ TeV. Hence there are no
values of $F$ for which there are solutions in {\it both} the low mass
and the high mass regions.  Furthermore, since the low mass region is
already excluded by LUX, the triplet case can only provide a DM
candidate for $F > 1.7$ TeV.

\subsubsection{Higher Representations}

Since the effective Lagrangian in eq.~(\ref{tripLeff0}) is valid for
any real DM representation, we can readily extrapolate the general
conclusions obtained for the triplet case to higher dimensional
representations. Since only odd representations contain a electrically
neutral component, the possible representations higher than the
triplet are the quintuplet ($n = 5$), the septuplet ($n =
7$)~\cite{Cirelli:2005uq}.  For any $n$-odd representation a mass
splitting between the neutral (DM) and charged components is always
induced by the dimension-6 operators proportional to $\lambda_{4,5}$
in composite scenarios.  We expect the allowed solutions in the low
mass region to be similar to the triplet case shown in
figure~\ref{fig:tripletLM}.  Consequently, in the case of higher
representations, most of the low mass region is expected to be
  excluded by experimental constraints or theoretically disfavored.

 For the high mass range, given the results obtained for the doublet
 and triplet DM, one can expect that the threshold mass is the same in
 composite and non-composite scenarios. In the latter case, one can
 check that the total effective annihilation cross-section in the
 early universe has a given dependence in
 $n$~\cite{Cirelli:2005uq,Hambye:2009pw}. In the pure gauge limit it
 scales as $(n^2-1)(n^2-3)/n$ for a multiplet of dimension $n$ so that
 for $n=5 \,(7)$ one would have $m_{DM}^{min}\simeq 4.3 \, (7.5) $
 TeV. Again, one expect that Sommerfeld corrections will push
 these thresholds to even higher masses, see \cite{Cirelli:2007xd} in
 the non-composite limit.

If we consider compositeness as a solution to the little hierarchy
problem, we must require $F \lesssim 3$ TeV in order to obtain a
fine-tuning of the order of 1\% or
  lower~\cite{Panico:2012uw}. Therefore, since $m_{DM} < F$, we
   see that the triplet is the highest DM representation
  allowed.  Although there is still a small range of low DM masses
  allowed for all representations (50 GeV $\lesssim m_{DM} \lesssim
  60$ GeV), these solutions usually require extremely small values of
  $\lambda_{\xi h}$, which are unlikely to be generated in realistic
  models, where one expects $\lambda_{\xi h} \sim
  \mathcal{O}(0.01-1)$.  Consequently we conclude that {\it Dark
    Matter representations higher than the triplet are highly
    disfavored in composite DM models}.

\section{Conclusions}\label{conclusions}

The nature of DM is one of the greatest conundrums of our time.  In
spite of the fact that DM constitutes 85\% of the total matter in the
Universe, it continues to evade direct experimental observation. A
reason for that may be connected to the fact that DM may be not
elementary. Having this in mind we investigate composite multiplets
that can arise in composite Higgs models and study the conditions
under which they can be a suitable DM candidate.

We considered a class of composite models where the only composite
states present in the low energy effective theory are the Higgs and
the DM, the first being a bi-doublet and the latter a multiplet of
$SO(4)$, both pNGBs of a spontaneous broken global symmetry of a new
unknown strongly coupled sector. We constructed and parametrized the
most general effective Lagrangian up to dimension-6 operators under
these general assumptions. We then checked if the different DM $SO(4)$
multiplet candidates could account for all the DM assuming that the
freeze-out mechanism is driving the relic abundance.  For different DM
representations under $SU(2)_L$ (or $SO(4)$) we derived the main DM
observables (relic abundance, spin-independent DM-nucleon scattering
cross-section, annihilation cross-sections) and imposed experimental
constraints from direct and indirect detection experiments as well as
LHC constraints from invisible Higgs decays.

In the generic singlet DM scenario (arbitrary cosets), the DM-nucleon scattering
cross-section can be significantly suppressed and can evade all current
experimental bounds for $m_{DM}  \gsim 100$ GeV (except for a small
discontinuity near the Higgs resonance), if $F \gtrsim 0.8$ TeV.
However, once a specific model is considered (with a given coset), it can
be severely constrained by the DM observables. As an example, we
discussed the case $\mG/\mH = SO(6)/SO(5)$.

The composite doublet DM models differ from the singlet one due to
tree level DM-gauge boson couplings and co-annihilations between DM
multiplet components. There are two viable mass regions, one below the
Higgs mass and one above $\sim$ 500 GeV.  Most of the points in
the low mass region are either excluded by data or theoretically
disfavored. For $m_{DM} \gsim 500$ GeV, the annihilation into gauge
bosons get suppressed enough to account for the DM relic abundance,
just as in the IDM scenario.  In the case $\mG/\mH = SO(6)/SO(4)
\times SO(2)$ presented in section~\ref{sec:SO6SO4SO2}, the
dimension-6 operators allow for solutions only if $F> 2$ TeV. The DM
mass is then constrained to the range $600 (550)$ GeV $\lsim m_{DM}
\lsim 0.8 (1.2)$ TeV for $F= 2.1 (2.4)$ TeV.

The composite triplet DM scenario is also mostly excluded by data
  or theoretical considerations in the low mass region, despite
allowing for larger mass splittings between neutral and charged
components than in the non-composite scenarios. The large mass region
will only be partially tested by Xenon1T. We also mention that while
in the non-composite triplet scenario a non-observation of a signal in
Xenon1T would provide an upper bound on $m_{DM}$, in the composite
case this upper bound could be easily avoided. Finally, we have also
examined higher representations and concluded they are highly
disfavored in composite Higgs models if one requires small fine-tuning
in the EW sector (or $F < 3$ TeV).
  
We have shown that DM in various representations in composite models
can reproduce the correct relic abundance and still be compatible with
limits on the Higgs invisible width and from the non-observation of DM
in direct and indirect detection experiments.  These models will be
further put to the test by future LHC, direct and indirect detection
data, either excluding them completely or revealing some exciting new
physics.

\begin{acknowledgments}
  We are grateful to Gabriele Ferretti and Eduardo Ponton for useful discussions.
  This work was supported by Funda\c{c}\~ao de Amparo \`a Pesquisa do
  Estado de S\~ao Paulo (FAPESP) and Conselho Nacional de Ci\^encia e
  Tecnologia (CNPq).  This work was also supported by the
  U.S. Department of Energy (DOE) under grant Contract Number
  DE-SC00012567. R.Z.F. thanks the Kavli Institute for Theoretical
  Physics in UC Santa Barbara for its hospitality, where part of this
  work was completed. This research was also supported in part by the
  National Science Foundation under Grant No. NSF PHY11-25915 and by
  the European Union FP7 ITN INVISIBLES (Marie Curie Actions,
  PITN-GA-2011-289442).  A.L. thanks HEPHY Vienna for its hospitality,
  where part of this work was completed. LLH is supported through an
  “FWO-Vlaanderen” post doctoral fellowship project number
  1271513. LLH also recognizes partial support from the Strategic
  Research Program “High Energy Physics” of the Vrije Universiteit
  Brussel and from the Belgian Federal Science Policy through the
  Interuniversity Attraction Pole P7/37 “Fundamental Interactions”.
\end{acknowledgments}

\appendix

\section{Dimension Six Operators for Composite DM}\label{appD6}

In order to compute the DM observables relevant for our results in
section~\ref{sec:results}, it is essential to identify the relevant
operators induced by the new strong sector. In
sections~\ref{sec:singletRes}, \ref{sec:doubletRes} and
\ref{sec:tripletRes} we presented the effective Lagrangians considered
in our analyses. Here we will derive them in detail and justify (when
necessary) our choice of operators.  First we consider the SM
dimension-6 operators. In addition to the ones listed in
$\mathcal{L}_6$ (eq.~\ref{eq:l0}) there are other SM dimension-6
operators that are in agreement with the assumptions discussed in
section \ref{compDM} but were not considered in our analysis.
Assuming minimally-coupled theories and a specific basis, the
following dimension-6 operators can be induced at tree-level by
integrating out heavy states with spin $\leqslant 1$~\cite{SILH:07,
  EffHiggsCorbett:13, EffHiggsContino:13, EffHiggsElias:1302,
  EffHiggsElias:1308, Alonso:2012}:
\begin{align} \label{f:treeLevelcHQ}
&\frac{ c_{H Q1} \, i}{F^2}  \left( H^\dagger\sigma^j {\overleftrightarrow D^\mu} H \right) \left(  \bar{Q}_L \sigma^j \gamma_\mu  Q_L \right),~~ \frac{ c_{H Q2} \, i}{F^2}  \left( H^\dagger  {\overleftrightarrow D^\mu} H \right) \left(  \bar{Q}_L \gamma_\mu  Q_L \right), \\ \label{f:treeLevelcHud}
& \frac{ c_{H u} \, i}{F^2}  \left( H^\dagger  {\overleftrightarrow D^\mu} H \right) \left(  \bar{u}_R \gamma_\mu  u_R \right),~~\frac{ c_{H d} \, i}{F^2}  \left( H^\dagger  {\overleftrightarrow D^\mu} H \right) \left(  \bar{d}_R \gamma_\mu  d_R \right), ~~ \frac{ c_{H ud} \, i}{F^2}  \left( H^{c \dagger}  {\overleftrightarrow D^\mu} H \right) \left(  \bar{u}_R \gamma_\mu d_R \right) \\ \label{f:treeLevelH}
& \frac{a_{W} \, ig}{M_\rho^2} \left (H^\dagger \sigma^j {\overleftrightarrow D^\mu} H\right) D^\nu W^j_{\mu \nu}, \\ \label{f:treeLevelNH} 
&  \frac{ a_{2W} \, g^2 }{  M_\rho^2 \, g_\rho^2 }(D^\mu W_{\mu\nu})^j (D_\beta W^{\beta \nu})^j, ~~~~ \frac{ a_{2B} \, g'^2 }{M_\rho^2 \, g_\rho^2 }(\partial^\mu B_{\mu\nu}) (\partial_\beta B^{\beta\nu}), ~~~~\frac{a_{2G} \, g_s^2  }{M_\rho^2 \,  g_\rho^2}(D^\mu G_{\mu\nu})^a (D_\beta G^{\beta\nu})^a,
\end{align}
where $H^\dagger \sigma^i \overleftrightarrow{D}^\mu H \equiv
H^\dagger \sigma^i D^\mu H - \left(D^\mu H \right)^\dagger \sigma^i
H$, $M_\rho \simeq g_{\rho} F $, and $g_{\rho}$ is the typical
coupling of the Higgs doublet and SM fermions to the heavy resonances
($ g_{\rho}\lesssim 4\pi$).  The coefficients $a_i$ and $c_i$ are
$\mathcal{O}(1)$ numbers.  The operators in (\ref{f:treeLevelH},
\ref{f:treeLevelNH}) have at least a suppression of $1/g_\rho^2$
compared with the operators in $\mathcal{L}_6$.\footnote{Note that the
  suppression of the operators in (\ref{f:treeLevelH},
  \ref{f:treeLevelNH}) compared with $\mathcal{L}_6$ is larger as the
  theory is close to the strong-coupling limit, i.e. when the typical
  coupling is $g_\rho\sim 4\pi$.}  Additionally, the operator
proportional to $a_W$ (\ref{f:treeLevelH}) contributes at tree-level
to the $S$ parameter, which is very constrained by the electroweak
precision data \cite{ SeTFit2}.  For these reasons, we do not consider
these operators in our effective Lagrangians. On the other hand,
(\ref{f:treeLevelcHQ}, \ref{f:treeLevelcHud}) have the same
suppression factor $(1/F^2)$ as the operators in $\mathcal{L}_6$ and
should, in principle, be considered. However, these operators modify
the gauge couplings to fermions which are in great agreement with the
SM \cite{SeTFit2}. We can try to avoid this constraint by absorbing
the universal part of the vertex corrections to a redefinition of the
electroweak gauge fields \cite{Agashe:2003, Burdman:2012}.  This
restores the gauge couplings to their SM values, but it generates
contributions to the electroweak precision parameters, which also are
severely constrained by the experimental data.  Therefore, all the
operators in (\ref{f:treeLevelcHQ}, \ref{f:treeLevelcHud}) are
extremely constrained and we disregard them.  We also neglect the
4-fermion operators since they do not affect the DM observables
discussed here. Finally, operators that can only be generated at
one-loop level, such as $\bar{Q}_L \sigma^j\sigma^{\mu\nu}\psi_R H^c
W_{\mu\nu}^j$, are suppressed with respect to the operators in
$\mathcal{L}_6$ and can also be neglected.

In the remaining sections we compute the $\xi$-dependent part of the
effective DM Lagrangian ($\mathcal{L}^{(2)}$) for specific DM
representations. In order to be consistent and also to reduce the
number of free parameters, we assume in the DM sector the same
simplifications previously required in the Higgs sector, i.e.
\begin{itemize}
\item we neglect the suppressed operators;
\item from the electroweak precision data \cite{SeTFit2} we know that the $c$'s  in (\ref{f:treeLevelcHQ}, \ref{f:treeLevelcHud}) must be very small $(c_{H \psi} \ll 1)$. Here we assume that the analogous coefficients in the DM sector are also suppressed.
\end{itemize}

\subsection{Singlet DM}\label{singletLag}

The case of singlet DM is trivial since $\xi$ can only couple to operators
which are singlet under $SU(2)_L \times U(1)_Y$.
The only operators (up to dimension 6) containing two DM
fields and the Higgs field contributing to the scalar potential are given by
\begin{align}
& \xi^2 \mbox{ , } \xi^2 |H|^2\mbox{ , } \xi^2 |H|^4.
\end{align}
The dimension-6 derivative interactions, corresponding to the effective CCWZ, must
preserve the symmetry $SO(4) \subset \mH$, hence the only allowed operators are
\begin{align}
\xi\dmu\xi \dMu |H|^2
 \mbox{ , } |H|^2 \dmu \xi \dMu \xi \mbox{ ,  } \xi^2 \left( \Dmu H \right)^\dagger \DMu H \,.
\end{align}
The last two operators can be eliminated after the following field redefinitions
\begin{equation}
H \to H + \frac{a}{4 F^2} \xi^2 H \mbox{ ,  }~~\xi \to \xi + \frac{a'}{F^2}
|H|^2 \xi \,,
\end{equation}
then,
\begin{align}
\left(\Dmu H\right)^\dagger \DMu H + \frac{1}{2} \dmu \xi \dMu \xi & \to
\left(\Dmu H\right)^\dagger \DMu H + \frac{1}{2} \dmu \xi \dMu \xi +
\frac{a + a'}{2 F^2} \xi\dmu\xi \dMu |H|^2 \nonumber
\\
& + \frac{a}{F^2} \xi^2\left(\Dmu H\right)^\dagger \Dmu H +
\frac{a'}{F^2} |H|^2 \dmu \xi \dMu \xi \nonumber +
\mathcal{O}\left(\frac{1}{F^4}\right)\,.
\end{align}
From the above result we see that choosing the coefficients $a$ and $a'$
properly, the operators $|H|^2 \dmu \xi \dMu \xi$ and
$\xi^2 \left(\Dmu H\right)^\dagger \DMu H$ can always be
eliminated.
Therefore, we obtain the following effective Lagrangian for the singlet DM field
\begin{align}
\mathcal{L}^{(2)} & = \frac{1}{2} \dmu \xi \dMu \xi
 - \frac{1}{2}\mu_{\xi}^2 \xi^2 - \frac{\lambda_3}{2}\left(1 +
 \frac{\lambda_3'}{F^2} |H|^2\right) \xi^2 |H|^2  + \frac{a_{d1}}{F^2} \dmu
 \xi^2 \dMu |H|^2    \nonumber\\
 & - \frac{1}{2} \frac{d_4}{F^2}\xi^2 \left(y_t \bar{Q}_L H^c t_R + y_b
 \bar{Q}_L H b_R  + \text{h.c.} \right) \,,\label{singLeffAppendix} 
\end{align}
where $d_4$ is real as we assume a $CP$-even Higgs.

\subsection{Doublet Dark Matter}\label{doubletLag}

The doublet DM, $\xi \sim \left(2, 2\right)$, is quite involved since in this
case $\xi$ has a non-zero hypercharge. However, the number of possible operators
can be greatly simplified imposing $SO(4)$ invariance and using suitable field redefinitions.
In order to simplify the notation, we use the following $SO(4)$ bi-doublets to construct the invariant operators
\be
\Phi \equiv \left(H^c, H\right)~~\mbox{ and }~~\Phi_{\xi} = \left(\xi^c,
\xi\right)\,.
\ee
The above $2\times2$ fields transform under $SO(4)$ as
\be
\Phi \to L\Phi R^\dagger~~\mbox{ and }~~\Phi_{\xi} \to L\Phi_{\xi} R^\dagger\,,
\ee
so $\text{Tr}\left[\Phi^\dagger \Phi \right]$ is a $SO(4)$ singlet.

For the scalar potential we can construct the following $SU(2)_L \times U(1)_Y$
invariant terms involving 2 powers of $\xi$
\begin{align}
 \label{f:DoubletDim2} \mbox{ dimension 2: } & \text{Tr}\left[\Phi_\xi^\dagger \Phi_\xi \right]\,; \\ \nonumber  ~\\ \nonumber
\mbox{ dimension 4: } & \text{Tr}\left[\Phi^\dagger \Phi \right] \, 
\text{Tr}\left[\Phi_\xi^\dagger \Phi_\xi \right] \mbox{ , } 
\text{Tr}\left[\Phi^\dagger \Phi_\xi \right]^2 \mbox{ , } \\ 
& \text{Tr}\left[\Phi^\dagger \Phi_\xi \sigma^3 \right]^2 \mbox{ , } 
\text{Tr}\left[\Phi^\dagger \Phi_\xi \right] \text{Tr}\left[\Phi^\dagger\Phi_\xi \sigma^3  \right]\,;  \label{f:DoubletDim4}  \\  \nonumber  \\
\mbox{ dimension 6: } & \text{Tr}\left[\Phi^\dagger \Phi \right]^2
\text{Tr}\left[\Phi_\xi^\dagger \Phi_\xi \right] \mbox{ , } 
\text{Tr}\left[\Phi^\dagger \Phi\right] \text{Tr}\left[\Phi^\dagger \Phi_\xi \right]^2\,,
\nonumber\\
& \text{Tr}\left[\Phi^\dagger \Phi \right]\text{Tr}\left[\Phi^\dagger \Phi_\xi \sigma^3 
\right]^2 \mbox{ , } \text{Tr} \left[\Phi^\dagger \Phi\right]\text{Tr}\left[\Phi^\dagger
\Phi_\xi \right] \text{Tr}\left[\Phi^\dagger\Phi_\xi \sigma^3  \right]\,.
 \label{f:DoubletDim6}
\end{align}
Notice that all the operators involving $\sigma^3$ explicitly break $SU(2)_R$. However,
they are allowed in the scalar potential since it is generated by the  Yukawa and gauge couplings between the SM and the strong sector,  which violate the $SO(4)$ symmetry.
All the other possible operators (including the
triplet-triplet contractions) can be reduced to the ones above. Writing the above operators in terms of $\xi$
and $H$ we obtain the following scalar potential
\begin{align}
V(H,\xi) & = \mu_{\xi}^2 |\xi|^2 + \lambda_3 \left(1 +
\frac{\lambda_3'}{F^2}|H|^2\right) |\xi|^2 |H|^2 + \lambda_4 \left(1 +
\frac{\lambda_4'}{F^2}|H|^2\right) |\xi^\dagger H|^2 \nonumber
\\
& +\frac{\lambda_5}{2}\left[ \left(1 + \frac{\lambda_5'}{F^2}|H|^2\right)
\left(\xi^\dagger H\right)^2 + \text{h.c.} \right]. \label{doubPot}
\end{align}

In the doublet case there are two dimension-6 Yukawa-type operators involving $\xi$, the singlet-singlet and triplet-triplet operators
\begin{align}
& V_{\text{Yuk}}(\Phi,\xi) = \frac{d_4}{F^2}|\xi|^2
\left(y_t \bar{Q}_L H^c t_R + y_b \bar{Q}_L H b_R  + \text{h.c.}\right) \nonumber \\
& + \frac{d_6}{F^2} 
  \xi^\dagger \vec{\sigma} \xi \left(y_t
 \bar{Q}_L \vec{\sigma} H^c t_R - y_b \bar{Q}_L \vec{\sigma} H b_R + \text{ h.c.}
 \right) \nonumber\\
& + \frac{d_6'}{F^2} 
  \left(y_b \xi^{c\dagger} \vec{\sigma} \xi \bar{Q}_L \vec{\sigma} H^c
b_R + y_t \xi^{\dagger} \vec{\sigma} \xi^c  \bar{Q}_L \vec{\sigma} H t_R
+\text{ h.c.}\right)\,.\label{doubYuk}
\end{align}
In order to avoid too many irrelevant parameters in the
potential we take $d_6' = d_6$, which we do not expect to affect any of our results.

Finally, the dimension-6 $SO(4)$ invariant derivatives operators, corresponding to the effective CCWZ Lagrangian, can be constructed from the operators listed in
(\ref{f:DoubletDim2}, \ref{f:DoubletDim4}, \ref{f:DoubletDim6})
\begin{align}
& \text{Tr}\left[\Phi^\dagger\Phi\right] \, \text{Tr}\left[\Dmu\Phi_\xi^\dagger\DMu\Phi_\xi\right]\,,~~
\text{Tr} \left[\Phi_\xi^\dagger\Phi_\xi\right]\text{Tr} \left[\Dmu\Phi^\dagger\DMu\Phi\right]\,,~~
\text{Tr} \left[\Phi^\dagger\Phi_\xi\right]\text{Tr} \left[\Dmu\Phi^\dagger\DMu\Phi_\xi\right]\,,
\nonumber \\
&
\text{Tr} \left[\Phi^\dagger\Dmu\Phi_\xi\right]^2\,,~~
\text{Tr} \left[\Phi_\xi^\dagger\Dmu\Phi\right]^2\,,~~
\text{Tr} \left[\Phi^\dagger\Dmu\Phi_\xi\right] \text{Tr} \left[\Phi_\xi^\dagger\DMu\Phi\right]\,,~~
\text{Tr} \left[\Phi^\dagger\Dmu\Phi\right] \text{Tr} \left[\Phi_\xi^\dagger\DMu\Phi_\xi\right]\,,
\nonumber \\
& \text{Tr} \left[\Phi^\dagger\Dmu\Phi_\xi\Phi_\xi^\dagger\DMu\Phi\right]
- \text{Tr} \left[\Phi^\dagger\Dmu\Phi\Phi_\xi^\dagger\DMu\Phi_\xi\right]
\label{eq:CCWZops}.
\end{align}

The first three operators can be rewritten in terms of the others if we
consider suitable field redefinitions
\begin{align}
H & \to H + a \, \text{Tr} \left[\Phi_\xi^\dagger \Phi_\xi \right] H + b \, \text{Tr} \left[\Phi_\xi^\dagger \Phi \right] \xi \,,
\nonumber \\
\xi & \to \xi + a' \, \text{Tr} \left[\Phi^\dagger \Phi \right] \xi + b \, \text{Tr} \left[\Phi^\dagger \Phi_\xi \right] H \,,
\end{align}
where $a$ and $a'$ are chosen to cancel the first two operators and $b$ is
chosen to cancel the third one.
Furthermore, we notice that
\begin{align}
\text{Tr}\left[\Phi^\dagger\Dmu\Phi_\xi\right]^2 -
\text{Tr} \left[\Phi_\xi^\dagger\Dmu\Phi\right]^2 = \dmu \text{Tr} \left[\Phi^\dagger
\Phi_\xi\right] \left( \text{Tr} \left[\Phi^\dagger\Dmu\Phi_\xi\right] -
\text{Tr} \left[\Phi_\xi^\dagger\Dmu\Phi\right]\right).
\end{align}
The above operator can also be eliminated through the field redefinition
\begin{align}
H \to H + c \, \text{Tr}\left[\Phi_\xi^\dagger \Phi\right] \xi  \mbox{ , } ~~
\xi \to \xi - c \, \text{Tr}\left[\Phi^\dagger \Phi_\xi\right] H\,,
\end{align}
with the coefficient $c$ suitably chosen. Therefore, after the above field transformations we obtain only $4$ independent operators, resulting in the following effective CCWZ Lagrangian \footnote{In
(\ref{doubCCWZ}) we have replaced the operator
\be \nonumber
\text{Tr}\left[\Phi^\dagger\Dmu\Phi_\xi\right]^2 +
\text{Tr}\left[\Phi_\xi^\dagger\Dmu\Phi\right]^2 ~~ \mbox{ by }~~
\left(\text{Tr}\left[\Phi^\dagger\Dmu\Phi_\xi\right] +
\text{Tr}\left[\Phi_\xi^\dagger\Dmu\Phi\right]\right)^2 \mbox{, }
\ee
since they only differ by the operator proportional to $a_{d2}$.}
\begin{align}
\mathcal{L}^{(2)}_{\text{CCWZ}} & =
 \left(\Dmu \xi\right)^\dagger \DMu \xi 
+  \left(\Dmu H\right)^\dagger \DMu H 
+
\frac{a_{d1}}{2F^2}\text{Tr} \left[\Phi^\dagger\Dmu\Phi\right]\text{Tr} \left[\Phi_\xi^\dagger\DMu\Phi_\xi\right]
\nonumber \\
&+\frac{a_{d2}}{F^2}\text{Tr} \left[\Phi^\dagger\Dmu\Phi_\xi\right]\text{Tr} \left[\Phi_\xi^\dagger\DMu\Phi\right]
+ \frac{a_{d3}}{F^2} \left( \text{Tr} \left[\Phi^\dagger\Dmu\Phi_\xi\right]
+ \text{Tr} \left[\Phi_\xi^\dagger\Dmu\Phi\right] \right)^2 \nonumber \\
& +\frac{a_{d4}}{F^2} \left(
\text{Tr} \left[\Phi^\dagger\Dmu\Phi_\xi\Phi_\xi^\dagger\DMu\Phi\right] - \text{Tr} \left[\Phi^\dagger\Dmu\Phi\Phi_\xi^\dagger\DMu\Phi_\xi\right] \right).
\label{doubCCWZ}
\end{align}

Finally, rewriting the operators in terms of the $H$ and $\xi$ fields and
combining (\ref{doubPot}), (\ref{doubYuk}) and (\ref{doubCCWZ}), we obtain the
effective Lagrangian for the doublet DM field
\begin{align}
\mathcal{L}^{(2)} & =
 \left(\Dmu \xi\right)^\dagger \DMu \xi 
-  \mu_{\xi}^2 |\xi|^2 - \lambda_3 \left(1 + \frac{\lambda_3'}{F^2}
|H|^2\right) |\xi|^2 |H|^2  - \lambda_4 \left(1 +
\frac{\lambda_4'}{F^2}|H|^2\right) |\xi^\dagger H|^2 \nonumber \\
& - \frac{\lambda_5}{2} \left(1 + \frac{\lambda_5'}{F^2}|H|^2\right)
\left[\left(\xi^\dagger H\right)^2 + \text{h.c.} \right] 
+\frac{a_{d1}}{2F^2}\dmu|H|^2\dMu|\xi|^2 \label{doubLeff} 
\\
& +\frac{a_{d2}}{F^2}\left(H^\dagger \Dmu \xi + \text{h.c.} \right)\left(\xi^\dagger
\DMu H + \text{h.c.} \right) + \frac{a_{d3}}{F^2} \left[ \dmu \left(\xi^\dagger H +
\text{h.c.} \right) \right]^2  \nonumber \\
& +\frac{a_{d4}}{F^2} \left[ \xi^\dagger \overleftrightarrow{D}_{\mu} \xi
H^\dagger \overleftrightarrow{D}^{\mu} H +  \xi^\dagger
\overleftrightarrow{D}_{\mu} \xi^C H^{C\dagger} \overleftrightarrow{D}^{\mu} H 
- \xi^\dagger \vec{\sigma}
 \overleftrightarrow{D}_{\mu} \xi H^\dagger \vec{\sigma} \overleftrightarrow{D}^{\mu} H
 + \text{h.c.} \right] \nonumber \\
 & - \frac{d_4}{F^2}
 |\xi|^2 \left(y_t \bar{Q}_L H^c t_R + y_b \bar{Q}_L
H b_R + \text{h.c.}\right)  \nonumber \\
 & - \frac{d_6}{F^2} 
  \left[\xi^\dagger \vec{\sigma} \xi \left(y_t
 \bar{Q}_L \vec{\sigma} H^c t_R - y_b \bar{Q}_L \vec{\sigma} H b_R \right) 
+  y_b \xi^{c\dagger} \vec{\sigma} \xi \bar{Q}_L \vec{\sigma} H^c b_R
+ y_t \xi^{\dagger} \vec{\sigma} \xi^c  \bar{Q}_L \vec{\sigma} H t_R + \text{h.c.}
\right]\,, \nonumber 
\end{align}
where  $d_4$ and $d_6$ are real as we assume that the  Higgs doublet is $CP$-even. We have verified that the imaginary parts of $\lambda_5$ and $\lambda_5'$  have almost no impact on the allowed
parameter space.  Therefore,  for simplicity, we take these complex parameters as real.

\subsection{Real Representations}
\label{tripletLag}

Here we discuss the case where $\xi$ is a real multiplet, which corresponds to the $SO(4)$ representations $(n,1)$. The possible operators involving a real
$\xi$ multiplet are greatly reduced due to the identities
\begin{equation}
\xi^\dagger \overleftrightarrow{D}^\mu \xi = \left(\Dmu \xi\right)^\dagger \vec{\Gamma} \xi + \text{h.c.} =
\xi^\dagger \vec{\Gamma} \xi = \left(\Dmu \xi\right)^\dagger \vec{\Gamma} \DMu
\xi =  0 \,. \label{zeros}
\end{equation}
Furthermore, since $\xi$ is a singlet under $SU(2)_R$, the $SO(4)$ invariance
implies that $\xi$ can only couple to $SU(2)_R$ singlets. The only
possible operators contributing to the scalar potential with two powers of $\xi$
are
\begin{align}
\mbox{ dimension 2: } & \text{Tr}\left[\Phi_\xi^\dagger \Phi_\xi\right]\,; \\ \nonumber ~ \\
\mbox{ dimension 4: } & \text{Tr} \left[\Phi^\dagger \Phi\right]\,;
\text{Tr} \left[\Phi_\xi^\dagger \Phi_\xi\right]
 \\ \nonumber ~\\
\mbox{ dimension 6: } & \text{Tr} \left[\Phi^\dagger \Phi\right]^2
\text{Tr} \left[\Phi_\xi^\dagger \Phi_\xi\right] \mbox{ , } \text{Tr} \left[\Phi^\dagger
\sigma^i \Phi \sigma^3\right] \text{Tr} \left[\Phi^\dagger
\sigma^j \Phi \sigma^3\right] \xi^\dagger \left\{\Gamma^i,\Gamma^j \right\} \xi
\mbox{ , } \nonumber \\
 & \text{Tr} \left[\Phi^\dagger
\sigma^i \Phi \sigma^+\right] \text{Tr} \left[\Phi^\dagger
\sigma^j \Phi \sigma^-\right] \xi^\dagger \left\{\Gamma^i,\Gamma^j \right\} \xi\,,
 \label{eq:VopsMult}
\end{align}
where $\sigma^\pm = \sigma^1 \pm i \sigma^2$.
The dimension-6 derivative operators invariant under $SO(4)$, which parametrize
the effective CCWZ Lagrangian, are given by
\begin{align}
 \text{Tr} \left[\Phi^\dagger\Phi\right] \left(\Dmu \xi\right)^\dagger \DMu\xi\,, ~~~
&  
|\xi|^2 \text{Tr} \left[\DMu \Phi^\dagger \Dmu \Phi\right]\,
\nonumber \\
\text{Tr}\left[\Phi^\dagger\Dmu\Phi\right] \dMu |\xi|^2
 & ~~\mbox{ and }~~ \text{Tr}\left[\Phi^\dagger \vec{\sigma} \Dmu \Phi \right] \xi^\dagger
\vec{\Gamma} \xi.
\label{eq:CCWZopsMult}
\end{align}
As before, the first two operators can be eliminated through the field
rescalings:
\begin{equation}
H \to H + a \, |\xi|^2 H \; \text{and} \; \xi \to \xi + a' \, |H|^2 \xi \,.
\end{equation}

Therefore, writing the above operators in terms of the $H$ and $\xi$ fields we
obtain
\begin{align}
\mathcal{L}^{(2)} & =  \left(\Dmu \xi\right)^\dagger \DMu \xi 
-  \mu_{\xi}^2 |\xi|^2 - \lambda_3 \left(1 + \frac{\lambda_3'}{F^2}
|H|^2\right) |\xi|^2 |H|^2 \nonumber \\
& - \frac{\lambda_4}{F^2} \xi^\dagger \left\{ \Gamma^i,\Gamma^j \right\} \xi
H^\dagger \sigma^i H H^\dagger \sigma^j H
- \frac{\lambda_5}{F^2} \xi^\dagger \left\{ \Gamma^i,\Gamma^j \right\} \xi
H^{c\dagger} \sigma^i H H^\dagger \sigma^j H^c \label{tripLeff}
 \\
& + \frac{a_{d1}}{2F^2}\dmu |\xi|^2 \dMu |H|^2
- \frac{a_{d4}}{F^2} \xi^\dagger \vec{\Gamma}
\overleftrightarrow{D}^\mu \xi H^\dagger \vec{\sigma} \overleftrightarrow{D}_\mu
H
\nonumber \\
 & - \frac{d_4}{F^2}|\xi|^2 \left(y_t \bar{Q}_L H^c t_R + y_b \bar{Q}_L H
 b_R  + \text{h.c.}\right) \nonumber,
\end{align}
where  once again we take $d_4$ to be real as we assume a $CP$-even Higgs.

\bibliography{bibcomp}{} \bibliographystyle{unsrt}
\end{document}